\documentclass[sigconf,natbib=true]{acmart}
\AtBeginDocument{%
  \providecommand\BibTeX{{%
    \normalfont B\kern-0.5em{\scshape i\kern-0.25em b}\kern-0.8em\TeX}}}

\copyrightyear{2023}
\acmYear{2023}
\setcopyright{acmlicensed}\acmConference[CIKM '23]{Proceedings of the 32nd ACM International Conference on Information and Knowledge Management}{October 21--25, 2023}{Birmingham, United Kingdom}
\acmBooktitle{Proceedings of the 32nd ACM International Conference on Information and Knowledge Management (CIKM '23), October 21--25, 2023, Birmingham, United Kingdom}
\acmPrice{15.00}
\acmDOI{10.1145/3583780.3614977}
\acmISBN{979-8-4007-0124-5/23/10}

\newcommand{\stitle}[1]{\vspace*{0.4em}\noindent{\bf #1.\/}}
\newcommand{\disgcn}{EDDA}
\newcommand{\name}{EDDA}

\usepackage{multirow}
\usepackage{graphicx}
\usepackage{float}
\usepackage{amsmath}
\usepackage{subfig}
\usepackage{caption}
\usepackage{algorithm}
\usepackage{algorithmicx}
\usepackage[noend]{algpseudocode}
\usepackage[normalem]{ulem}
\usepackage{longtable}
\usepackage{pifont}
\useunder{\uline}{\ul}{}

\begin{document}

\title{Multi-domain Recommendation with Embedding Disentangling and Domain Alignment}


\author{Wentao Ning}
\affiliation{
    \institution{The University of Hong Kong}
    \institution{Southern University of Science and Technology}
    \country{}
}
\email{nwt9981@connect.hku.hk}

\author{Xiao Yan}
\affiliation{
    \institution{Southern University of Science and Technology}
    \country{}
}
\email{yanx@sustech.edu.cn}

\author{Weiwen Liu}
\affiliation{
    \institution{Huawei Noah’s Ark Lab}
    \country{}
}
\email{liuweiwen8@huawei.com}

\author{Reynold Cheng}
\authornote{Reynold Cheng is a corresponding author. He is also affiliated with Department of Computer Science and Musketeers Foundation Institute of Data Science, The University of Hong Kong and Guangdong–Hong Kong-Macau Joint Laboratory.}
\affiliation{
    \institution{The University of Hong Kong}
    \country{}
}
\email{ckcheng@cs.hku.hk}

\author{Rui Zhang}
\affiliation{
    \institution{www.ruizhang.info}
    \country{}
}
\email{rayteam@yeah.net}

\author{Bo Tang}
\authornote{Bo Tang is a corresponding author. He is also affiliated with Research Institute of Trustworthy Autonomous Systems and Department of Computer Science and Engineering, Southern University of Science and Technology, Shenzhen, China.}
\affiliation{
    \institution{Southern University of Science and Technology}
    \country{}
}
\email{tangb3@sustech.edu.cn}

\renewcommand{\shortauthors}{Wentao Ning et al.}


\begin{abstract} 

Multi-domain recommendation (MDR) aims to provide recommendations for different domains (e.g., types of products) with overlapping users/items and is common for platforms such as Amazon, Facebook, and LinkedIn that host multiple services. Existing MDR models face two challenges: First, it is difficult to disentangle knowledge that generalizes across domains (e.g., a user likes cheap items) and knowledge specific to a single domain (e.g., a user likes blue clothing but not blue cars). Second, they have limited ability to transfer knowledge across domains with small overlaps. We propose a new MDR method named \name~ with two key components, i.e., \textit{embedding disentangling recommender} and \textit{domain alignment}, to tackle the two challenges respectively. In particular, the embedding disentangling recommender separates both the model and embedding for the inter-domain part and the intra-domain part, while most existing MDR methods only focus on model-level disentangling. The domain alignment leverages random walks from graph processing to identify similar user/item pairs from different domains and encourages similar user/item pairs to have similar embeddings, enhancing knowledge transfer. We compare \name~ with 12 state-of-the-art baselines on 3 real datasets. The results show that \name~ consistently outperforms the baselines on all datasets and domains. All datasets and codes are available at \url{https://github.com/Stevenn9981/EDDA}.

\end{abstract}

\ccsdesc[500]{Information systems~Recommender systems}


\keywords{multi-domain recommendation; disentanglement; alignment}


\maketitle

\section{Introduction}\label{sec:intro}


\textit{Multi-domain recommendation} (MDR)~\cite{CPT, MGE, ADIN, EGSMDR} refers to the task of providing recommendations for multiple domains or scenarios where there are overlapping users and items. 
MDR is applicable in various contexts, such as e-commerce platforms like Amazon, where different domains of commodities (e.g., vehicles, laptops, fitness tools) are recommended to users, and users may interact with multiple domains.
Similarly, an item can belong to multiple domains, such as bicycles that can be in both [vehicles] and [fitness tools] domains. Other examples of MDR include Facebook recommends posts, user groups and videos, WeChat recommends official accounts, videos and mini-programs, and LinkedIn recommends jobs, pages and people. 
The overlapping users and items among different domains provide an opportunity to leverage users' interests in one domain to enhance recommendations in other domains. 

\begin{figure}[]
	\centering
	\includegraphics[width=0.8\columnwidth]{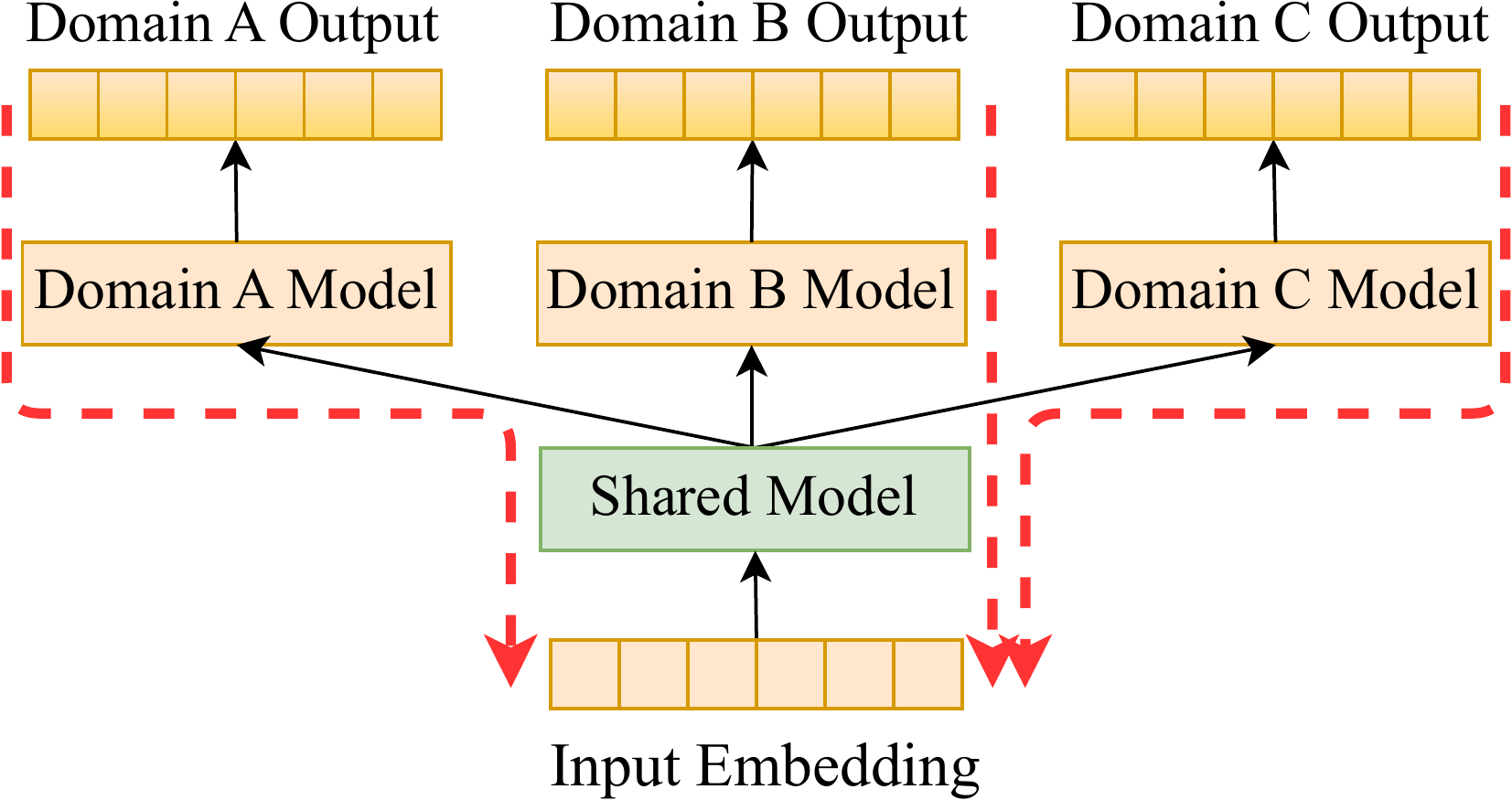}
	\caption{The model disentangling architecture in existing MDR methods. Shared model learns the inter-domain knowledge that generalizes across domains. Domain-specific models learn intra-domain knowledge that only works for each domain. The red dashed line represents the back-propagation flow. Both intra-domain and inter-domain knowledge will be used to update the input embedding.}
	\label{fig:model-level}
	\vspace{-2em}
\end{figure}


However, we find two key challenges in MDR. 
\ding{182} It is difficult to exploit useful information from other domains while avoiding the negative impact on recommendation accuracy~\cite{WangZJLBMZZ21,LiuLG19,SatoNHMHAM18}.
For example, a user may prefer cheap commodities, which is a common property that can guide recommendations in all domains. In contrast, a user may favor Apple's laptops in the [Laptops] domain because her previous laptop is Apple, yet this does not indicate that she will like Apple's loudspeakers in the [Speaker] domain. Generally speaking, there is \textit{inter-domain knowledge} that generalizes across domains and \textit{intra-domain knowledge} that only works for a specific domain, and distinguishing them is crucial for accurate recommendations.
\ding{183} Effective knowledge sharing among different domains is challenging, particularly when the overlap of users and items between domains is small (i.e., \textit{overlap sparsity}) because these common users/items usually serve as the bridge for knowledge sharing. In addition, for domains with limited user-item interaction data (i.e., \textit{data sparsity}), knowledge sharing is more critical to recommendation accuracy because small domains usually contain less information and external knowledge can help avoid over-fitting~\cite{BPAM, CIGAR} and bring more benefits.




Most existing MDR approaches~\cite{StarShengZZDDLYLZDZ21, TreeMSNiuLLT0D21, MGFNZhangPWPZLQ22} conduct model-level disentangling (as illustrated in Figure~\ref{fig:model-level}, referred to as \textit{model disentangling}) and focus on designing better domain-shared and domain-specific models. 
For example, 
STAR~\cite{StarShengZZDDLYLZDZ21} leverages fully-connected networks (FCNs) as domain-shared and domain-specific models. 
TreeMS utilizes GAT~\cite{GATVelickovicCCRLB18} as the shared model, and combines FCN and LSTM~\cite{LSTM} in domain-specific model design.
MGFN~\cite{MGFNZhangPWPZLQ22} uses GAT~\cite{GATVelickovicCCRLB18} to learn both intra-domain and inter-domain knowledge.
However, the problem with model disentangling is that inter-domain and intra-domain knowledge are still entangled in a single embedding for each user/item. The inter-domain knowledge and each specific domain's knowledge may lead to different gradient update directions. Since gradients are accumulated, they may cancel each other out if they point in opposite directions and then result in sub-optimal performance~\cite{CAGrad, YuK0LHF20, RotoGrad}.
Thus, model disentangling suffers from the \textit{gradient conflict} problem and cannot address it well. 
In addition, existing methods lack an effective mechanism to transfer knowledge across domains because they realize knowledge sharing only relying on the overlapping users/items among domains~\cite{CFAA, MGFNZhangPWPZLQ22, LiuLLP20}. These approaches face limitations, especially when the overlapping users/items are limited.


In this paper, we propose a new method \name~ for MDR, which consists of an \textit{embedding disentangling (ED) recommender} and a \textit{domain alignment} strategy. \textit{Firstly},
The ED recommender can exploit knowledge from other domains while avoiding negative transfer (challenge \ding{182}). It explicitly disentangles inter-domain and intra-domain knowledge into separate embeddings for each user/item. For every user/item, \name~ trains both an \textit{inter-domain embedding} that captures knowledge across all domains and an \textit{intra-domain embedding for each domain}. These embeddings are projected by the respective inter-domain and intra-domain models and then concatenated to conduct recommendations for each domain. Unlike model disentangling, ED recommender explicitly separates inter-domain and intra-domain knowledge into distinct embeddings, effectively avoiding the gradient conflict problem.
\textit{Secondly}, to enhance knowledge sharing across domains and alleviate the overlap and data sparsity problem (challenge \ding{183}), we propose a \textit{domain alignment} strategy based on random walks to identify similar user/item pairs from different domains. Specifically, we use the overlapping users/items in two domains as anchors and perform random walks~\cite{LiSDCC22, XiongY22} on the user-item interaction graph of each domain. For each user/item, we calculate its random walk stopping probability on the anchors, which is used to define the similarity between users/items from different domains. The rationale is that two users/items should exhibit similar behavior patterns or properties if they are likely to reach the same set of anchors. We achieve knowledge sharing by encouraging similar users/items from different domains to have similar embeddings.

We conduct extensive experiments to compare \name~ with 12 state-of-the-art baselines on 3 real datasets each containing 3 to 8 domains. The results show that \name~ significantly outperforms the baselines for all datasets and all domains. In particular, the improvements of \name~ over the best-performing baseline in AUC and recall are up to 7.6\% and 41.8\%, respectively. 

To sum up, we make the following contributions in this paper:

\begin{itemize}
    \item Propose an \textit{embedding disentangling recommender} for MDR, which explicitly disentangles inter-domain and intra-domain knowledge at both model and embedding level.
    \item Propose a random walk-based \textit{domain alignment} strategy to identify similar users/items from different domains, which further helps knowledge sharing.
    \item Conduct extensive experiments to evaluate \name~ that integrates the above two components, demonstrating its effectiveness, and validating the key designs.
\end{itemize}

The rest of the paper is organized as follows. Section~\ref{sec:pr} gives the definition of multi-domain recommendation. Section~\ref{sec:disgcn} presents our embedding disentangling recommender, and Section~\ref{sec:rw} elaborates on our random walk-based domain alignment strategy. Section~\ref{sec:exp} reports the experimental results. Section~\ref{sec:rel} discusses the related works, and Section~\ref{sec:con} draws the concluding remarks.

\section{Problem Definition} \label{sec:pr}

\stitle{Multi-domain recommendation (MDR)} 
A domain $\mathbf{d}$ can be represented as $\mathbf{d}=\{\mathcal{U}^\mathbf{d}, \mathcal{I}^\mathbf{d}, \mathcal{R}^\mathbf{d}\}$, where $\mathcal{U}^\mathbf{d}$ is the user set, $\mathcal{I}^\mathbf{d}$ is the item set, and $\mathcal{R}^\mathbf{d}$ is the set of user-item interactions (e.g., clicks, purchases). MDR deals with a set of domains $\mathbf{D} = \{\mathbf{d}_1, \mathbf{d}_2,\cdots, \mathbf{d}_W\}$, where $\mathbf{d}_w$ refers to a specific domain and $W$ is the number of domains. There may exist some common users and items for two different domains $\mathbf{d}$ and $\mathbf{d}'$ (called overlap), i.e., $\mathcal{U}^\mathbf{d}\cap \mathcal{U}^{\mathbf{d}'}\neq \emptyset$ or $\mathcal{I}^\mathbf{d}\cap \mathcal{I}^{\mathbf{d}'}\neq \emptyset$, which allows to share knowledge across domains and achieve better recommendation accuracy than handling each domain individually. The task is to find the items that a user is likely to interact with for each domain $\mathbf{d}_w \in \mathbf{D}$. Specifically, we want to learn a score function $f(u, i| {\mathbf{d}_w})$ that estimates the preference of user $u$ for item $i$, with which a set of items with the highest scores can be recommended to each user. 


\section{ED Recommender}
\label{sec:disgcn}

\begin{figure*}[]
	\centering
	\includegraphics[width=0.85\textwidth]{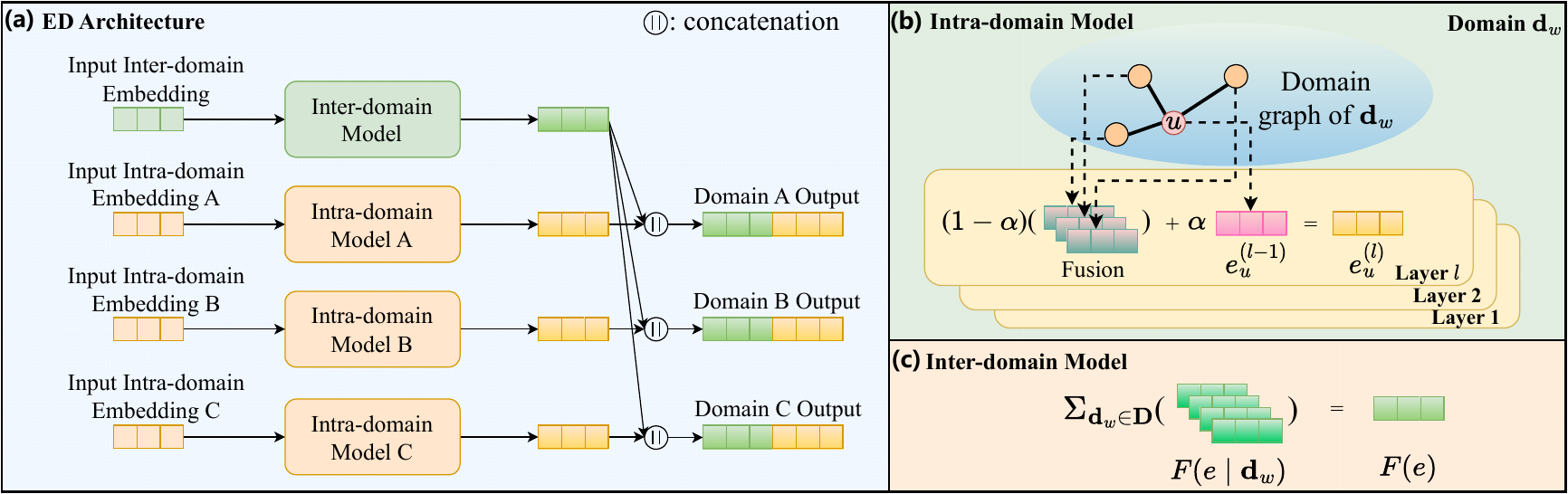}
	\caption{Illustration of ED recommender. The trainable parameters are input embeddings and model parameters.}
	\label{fig:framework}
	\vspace{-1em}
\end{figure*}

To effectively complement the domains with each other and avoid negative transfer, we propose the embedding disentangling (ED) recommender, as depicted in Figure~\ref{fig:framework}. The ED architecture shown in Figure~\ref{fig:framework}(a) is an overview of our ED recommender, and the implementations of intra-domain and inter-domain models are shown in Figure~\ref{fig:framework}(b) and (c), respectively.
In this section, we first elaborate on our ED architecture and then illustrate the designs of our intra-domain and inter-domain models.

\subsection{ED Architecture}\label{subsec: eda}

Figure~\ref{fig:framework}(a) shows the ED architecture, which is an overview of our ED recommender. In the ED architecture, we utilize separate \textit{intra-domain embeddings} and \textit{inter-domain embeddings} for each user/item, which are learned through the \textit{intra-domain model} and \textit{inter-domain model}, respectively. The underlying purpose is to allow the intra-domain embeddings (and model) to capture domain-specific information while the inter-domain embeddings (and model) learn general information that is applicable across all domains.


Formally, each user $u$ (resp. item $i$) has an inter-domain embedding $\boldsymbol{e}_u$ (resp. $\boldsymbol{e}_i$) that works for all domains and an intra-domain embedding $\boldsymbol{e}_u^{\mathbf{d}_w}$ (resp. $\boldsymbol{e}_i^{\mathbf{d}_w}$) for each domain $\mathbf{d}_w$ she belongs to. Note that for each user/item, the intra-domain embeddings for different domains are different and initialized separately. We also have an inter-domain model $\boldsymbol{F}$ and a separate intra-domain model $\boldsymbol{F}^{\mathbf{d}_w}$ for each domain. To obtain the user/item representations for a domain $\mathbf{d}_w$, the inter-domain and intra-domain embeddings are projected to latent spaces by the inter-domain and intra-domain models and concatenated as follows.


\begin{equation}
    \begin{aligned}
        \boldsymbol{Z}^{\mathbf{d}_w}_u &= \boldsymbol{F}(\boldsymbol{e}_u) \  || \  \boldsymbol{F}^{\mathbf{d}_w}(\boldsymbol{e}^{\mathbf{d}_w}_u), \\
        \boldsymbol{Z}^{\mathbf{d}_w}_i &= \boldsymbol{F}(\boldsymbol{e}_i) \  || \ \boldsymbol{F}^{\mathbf{d}_w}(\boldsymbol{e}^{\mathbf{d}_w}_i),
    \end{aligned}
    \label{eq: 1}
\end{equation}
where $||$ denotes vector concatenation, $\boldsymbol{Z}^{\mathbf{d}_w}_u$ and $\boldsymbol{Z}_i^{\mathbf{d}_w}$ are the projected user and item representations for domain $\mathbf{d}_w$. With the projected user/item representations, we use inner product as the score function $f(u, i| {\mathbf{d}_w})$, which is widely used in recommendation tasks~\cite{0001DWLZ020, abs-1205-2618, RMS}. That is,


\begin{equation}
    f(u, i| {\mathbf{d}_w}) = {\boldsymbol{Z}_{u}^{\mathbf{d}_w}}^{\top} \boldsymbol{Z}_{i}^{\mathbf{d}_w}.
\end{equation} 

\stitle{Training} 
To train the (inter-domain and intra-domain) models and randomly initialized (inter-domain and intra-domain) embeddings, we use the Bayesian Personalized Ranking (BPR) ~\cite{abs-1205-2618} loss in Eq.(\ref{eq: cf}), which is popular for training recommendation models~\cite{abs-1205-2618, 0001DWLZ020, KGATWang00LC19}.
\begin{small}
\begin{equation} \label{eq: cf}	
		    L_{\text{BPR}} = \sum_{{\mathbf{d}_w} \in\mathbf{D}} \sum_{\left(u, i^{+}, i^{-}\right) \in \mathcal{O}^{\mathbf{d}_w}}-\ln \sigma\left(f(u, i^{+}| {\mathbf{d}_w})-f(u, i^{-}| {\mathbf{d}_w})\right),	
\end{equation}
\end{small}
where $\mathcal{O}^{\mathbf{d}_w}=\{(u, i^{+}, i^{-}) | \mathbf{d}_w \}$ denotes the training sample set for domain ${\mathbf{d}_w}$ and $(u, i^{+}, i^{-})$ is a triplet. In particular, $(u, i^{+})$ is a user-item pair that belongs to the observed user-item interactions $\mathcal{R}^{\mathbf{d}_w}$ of domain $\mathbf{d}_w$ while $(u, i^{-})$ is a randomly sampled unobserved user-item pair in domain $\mathbf{d}_w$. Thus, BPR loss encourages assigning higher scores to the observed user-item interactions than unobserved ones. 

\stitle{Discussion} 
From the perspective of gradient flow, our ED architecture completely decouples the inter-domain part (i.e., both model and embedding) and the intra-domain part, and allows them to learn independently without interfering each other. In particular, for each training triplet $(u, i^{+}, i^{-})$, gradients first propagate to the score function (i.e., model output), then update the inter-domain model and the intra-domain model, and finally update the inter-domain and intra-domain embeddings. Consider an item $i$, triplets related to $i$ in all domains can update its inter-domain domain embedding $\boldsymbol{e}_i$ while only triplets related to $i$ in domain ${\mathbf{d}_w}$ can update its intra-domain embedding $\boldsymbol{e}_i^{\mathbf{d}_w}$ for domain ${\mathbf{d}_w}$. Thus, our design goal is achieved---the inter-domain domain embedding learns knowledge for all domains while the intra-domain embedding learns for each individual domain.

In comparison, the model disentangling architecture in Figure~\ref{fig:model-level}, which is commonly adopted by existing works~\cite{StarShengZZDDLYLZDZ21, MMoEMaZYCHC18, MGFNZhangPWPZLQ22}, differentiates inter-domain and intra-domain models but uses a single embedding each for user and item in all domains. Their design entangles domain-specific and domain-general knowledge and causes interference among them (i.e., they may require updating the embedding in different directions and cause gradient conflict problems).  


\begin{figure}[t]
	\centering
	\includegraphics[width=0.7\linewidth]{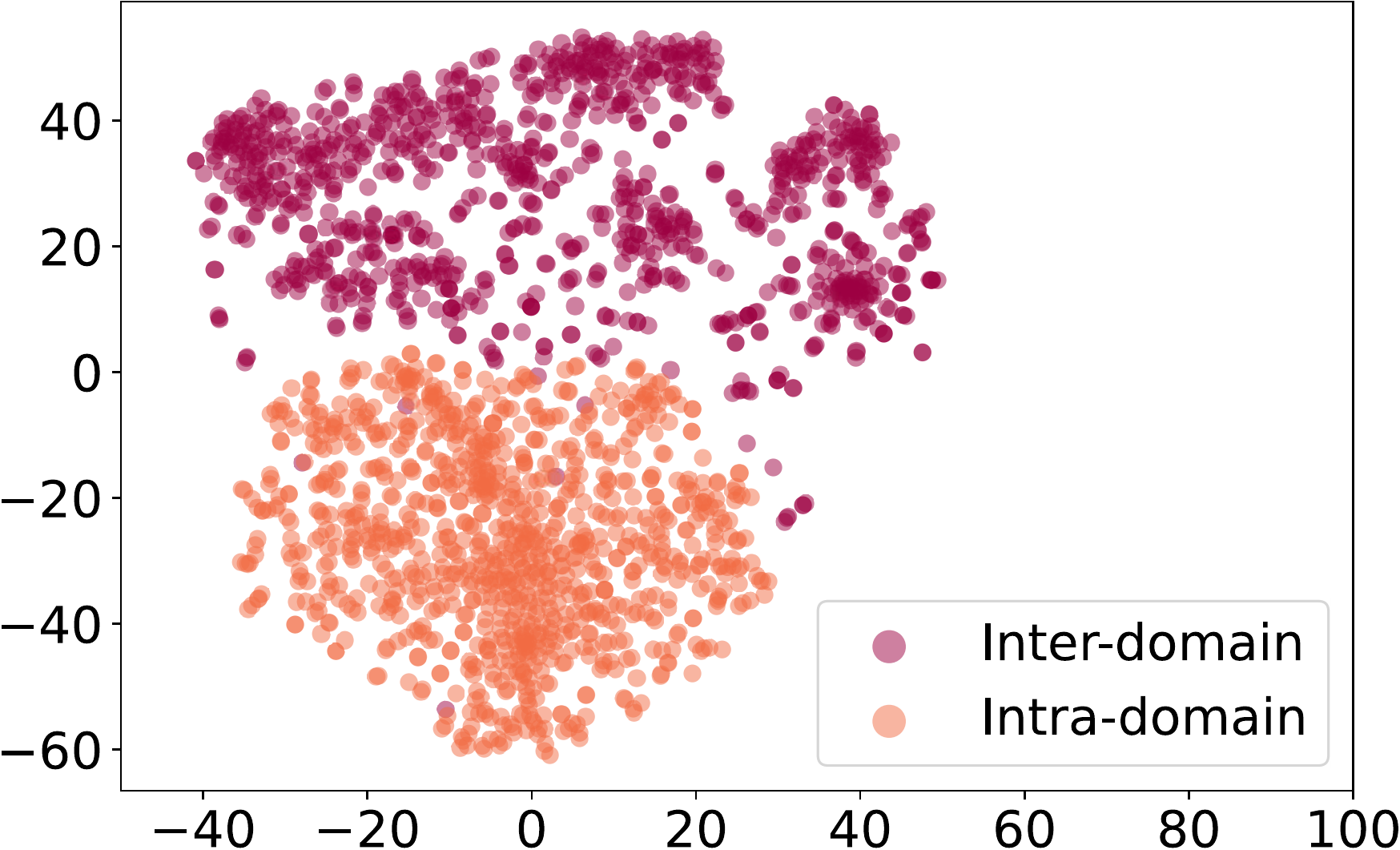}
	\caption{t-SNE visualization of embedding disentangling.}
	\label{fig:emb_vis}
	\vspace{-1.5em}
\end{figure}

To demonstrate that inter-domain and intra-domain embeddings learn different knowledge, we randomly sample 1000 items from one of our experiment datasets (domain D1 of AliAd, see the details in Section~\ref{subsec:expset}). Figure~\ref{fig:emb_vis} visualizes their inter-domain and intra-domain embeddings by reducing the embeddings to 2 dimensions using t-SNE~\cite{tsne}. The results show that the intra-domain and inter-domain embeddings form two well-separated clusters, suggesting that intra-domain and inter-domain knowledge are well disentangled. 






 

\subsection{Intra-domain and Inter-domain Models}\label{subsec: il}


\stitle{Intra-domain model} 
Consider a single domain ${\mathbf{d}_w}$, the user-item interactions can be represented as a bipartite graph $\mathcal{G}^{\mathbf{d}_w} = \{ \mathcal{V}^{\mathbf{d}_w}, \mathcal{E}^{\mathbf{d}_w} \}$. In particular, the node set $\mathcal{V}^{\mathbf{d}_w}=\mathcal{U}^{\mathbf{d}_w}\cup \mathcal{I}^{\mathbf{d}_w}$ comprises the users and items of the domain, while each edge in the edge set $\mathcal{E}^{\mathbf{d}_w}$ corresponds to a user-item interaction record in $\mathcal{R}^{\mathbf{d}_w}$. We refer to $\mathcal{G}^{\mathbf{d}_w}$ as the \textit{domain graph} of ${\mathbf{d}_w}$ since it encapsulates the data within domain ${\mathbf{d}_w}$. Here, we design a graph-based network, called \textit{GRec}, as our intra-domain model shown in Figure~\ref{fig:framework}(b). GRec learns user and item embeddings by performing iterative neighbor aggregation~\cite{Rui4AAAI21, Rui3VLDB22}, which captures the high-order connectivity of users and items in the domain graph. The detailed process is implemented using the following equations:

\begin{equation}
    \begin{aligned}
    &\boldsymbol{e}_u^{(l)}=\alpha \boldsymbol{e}_u^{(l-1)} + (1-\alpha) \sum_{i \in \mathcal{N}_u} \frac{1}{\sqrt{\left|\mathcal{N}_u\right|} \sqrt{\left|\mathcal{N}_i\right|}} \boldsymbol{e}_i^{(l-1)}, \\
    &\boldsymbol{e}_i^{(l)}=\alpha \boldsymbol{e}_i^{(l-1)} + (1-\alpha)\sum_{u \in \mathcal{N}_i} \frac{1}{\sqrt{\left|\mathcal{N}_u\right|} \sqrt{\left|\mathcal{N}_i\right|}} \boldsymbol{e}_u^{(l-1)},
    \end{aligned}
    \label{eq: msg}
\end{equation}

where $\boldsymbol{e}^{(l)}_u$ and $\boldsymbol{e}^{(l)}_i$ denote the embeddings of user $u$ and item $i$ in the $l$-th layer, and there can be $L$ layers. $\mathcal{N}_u$ and $\mathcal{N}_i$ are the set of neighbors of $u$ and $i$, respectively, and  $\sqrt{\left|\mathcal{N}_u\right|} \sqrt{\left|\mathcal{N}_i\right|}$ is a normalization term to adjust the magnitude of the embeddings during neighbor aggregation~\cite{Rui1SIGIR21, Rui2KDD22}. Inspired by~\cite{APPNP}, we introduce a hyper-parameter $\alpha$ that belongs to the range [0, 1] and controls the amount of information about the node itself that is retained during neighbor aggregation. We remove all linear projection functions and non-linear activation functions in traditional graph convolution layers~\cite{KipfW16}, as existing works have shown them to be harmful to capture collaborative filtering signals~\cite{0001DWLZ020, SGC}. Therefore, GRec has no other trainable parameters but the input user/item embeddings $\boldsymbol{e}^{(0)}_u$ and $\boldsymbol{e}^{(0)}_i$ (i.e., the 0-th layer embeddings). 

The goal of the intra-domain model is to learn domain-specific knowledge for a domain ${\mathbf{d}_w}$. Thus, we use a $L$-layer GRec and perform neighbor aggregation following Eq.~\eqref{eq: msg} only on the domain graph $\mathcal{G}^{\mathbf{d}_w}$ of ${\mathbf{d}_w}$.  The output of the last layer is used as the projected intra-domain embeddings $\boldsymbol{F}^{\mathbf{d}_w}(\boldsymbol{e}^{\mathbf{d}_w}_u)$ for each user/item.  The input intra-domain embedding for each user $u$ (item $i$) becomes the 0-th layer embeddings, i.e., $\boldsymbol{e}^{(0)}_u = \boldsymbol{e}^{\mathbf{d}_w}_u$ ($\boldsymbol{e}^{(0)}_i = \boldsymbol{e}^{\mathbf{d}_w}_i$), and we randomly initialize the input intra-domain embeddings.

\stitle{Inter-domain model} 
We also use an $L$-layer GRec as the inter-domain model but with different input user/item embeddings (i.e., \textit{inter-domain embeddings}). As the inter-domain model learns general knowledge across all domains, we apply GRec on the domain graphs from all domains and sum up the last layer embedding of a node in all graphs as the projected inter-domain embedding $\boldsymbol{F}(\boldsymbol{e}_u)$ as shown in Figure~\ref{fig:framework}(c).

\begin{equation}
    \begin{aligned}
        \boldsymbol{F}(\boldsymbol{E}_u) &=\!\sum_{{\mathbf{d}_w} \in \mathcal{D}, u\in \mathbf{d}_w} \!\!\!\! \boldsymbol{F}(\boldsymbol{E}_u|\mathbf{d}_w), \\
        \boldsymbol{F}(\boldsymbol{E}_i) &=\!\sum_{{\mathbf{d}_w} \in \mathcal{D}, i\in \mathbf{d}_w} \!\!\!\! \boldsymbol{F}(\boldsymbol{E}_i|\mathbf{d}_w),
    \end{aligned}
\end{equation}
where $\boldsymbol{E}_u$ and $\boldsymbol{E}_i$ are the randomly initialized input inter-domain embeddings of user $u$ and item $i$. $\boldsymbol{F}(\boldsymbol{E}_u|\mathbf{d}_w)$ denotes applying GRec on the domain graph of ${\mathbf{d}_w}$ using $\boldsymbol{E}_u$ as the 0-th layer embedding. 

It is worth noting that the input inter-domain embeddings are initialized independently from the intra-domain embeddings. Also, for each user/item, the same input inter-domain embedding is used for neighbor aggregation on all domain graphs in the inter-domain model.
Moreover, supervision signals can guide disentangling because the inter-domain embedding is updated by gradients from all domains and thus reflects inter-domain knowledge, while the intra-domain embedding is only updated by gradients from a specific domain and thus encodes domain-specific knowledge. 

Furthermore, GRec can be replaced by other existing single-domain recommendation models (e.g., MF~\cite{abs-1205-2618}) to make them suitable for MDR. Our experiments in Section~\ref{subsec:abl} demonstrate that using MF in the ED architecture as intra-domain and inter-domain models yields superior performance compared to training separate MF models for each domain individually.




\section{Domain Alignment}\label{sec:rw}

\begin{figure}[]
    \centering
    \includegraphics[width=7.5cm]{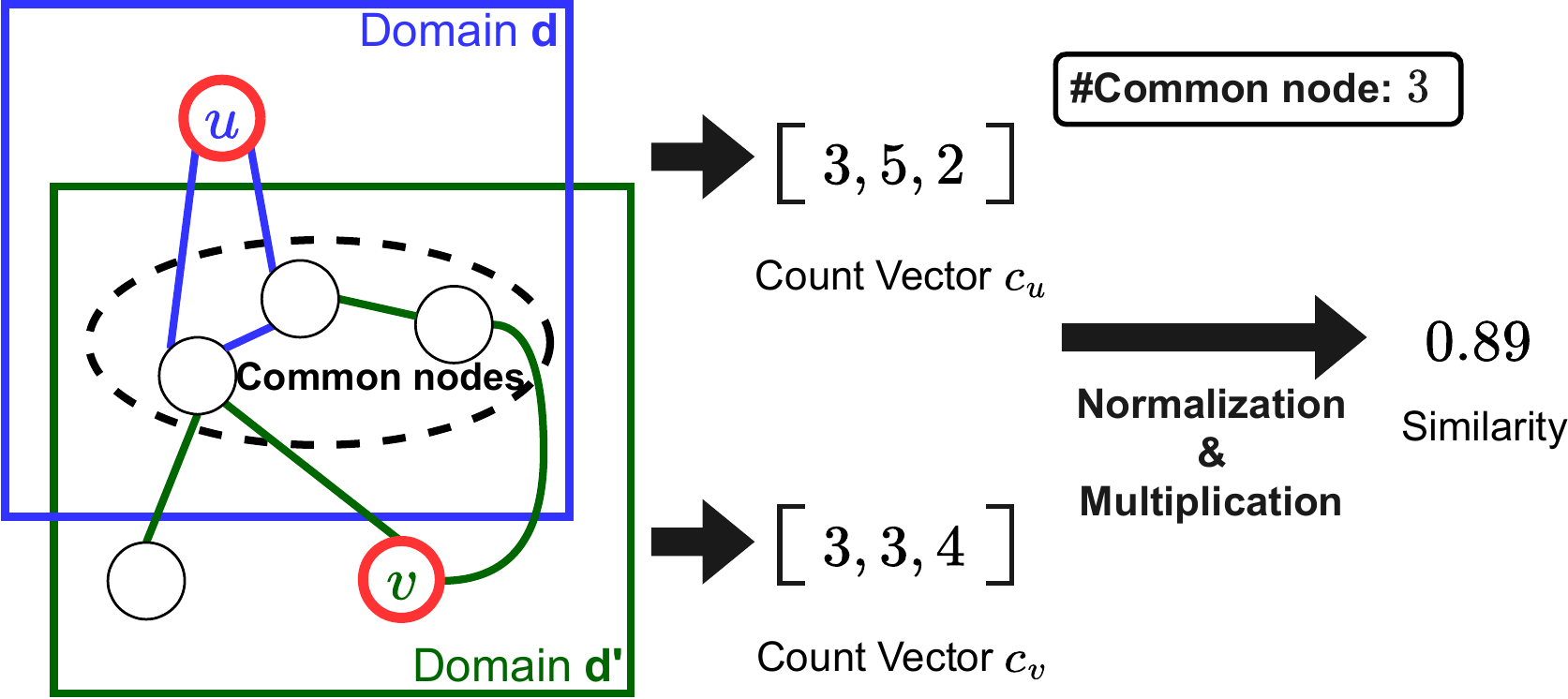}
    \caption{Illustration of our random walk-based node similarity. The similarity between the nodes from two domain graphs are computed using the random walk stop count vector on the common nodes of the two domains. There are 3 common nodes between domain ${\mathbf{d}}$ and ${\mathbf{d}'}$, and we perform 10 random walks from node $u$ and $v$ in this toy example. 
    Each entry of the count vector represents the number of times stopping at each common node.
    }
    \label{fig: rw}
    \vspace{-2em}
\end{figure}

For MDR, the overlapping users/items that appear in multiple domains play a crucial role in cross-domain knowledge transfer. However, in practice, the overlap between domains can be small. As shown in Table~\ref{tab:stat}, for the AliCCP dataset, the average overlap ratio between domains is only 0.019, and for the domain with the least overlap, the ratio is below 0.005. The limited overlap between domains restricts knowledge transfer since the common users/items act as the bridge for information sharing. 

Nevertheless, different users in different domains may exhibit similar behavior patterns, and items in different domains may share similar interaction patterns or properties since they may be preferred by users with similar interests. Thus, we propose a random walk-based domain alignment strategy that identifies similar user/item pairs in different domains. Then we let similar users/items have closer intra-domain embeddings to enhance knowledge transfer since they exhibit similarities in different domains. 








Consider two domains $\mathbf{d}$ and $\mathbf{d}'$ with domain graphs $\mathcal{G}^{\mathbf{d}}$ and $\mathcal{G}^{\mathbf{d}'}$, and assume set $\mathcal{T}$ contains common nodes (i.e., the overlapped users/items) that appear in both domain graphs. As shown in Figure~\ref{fig: rw}, our random walk-based procedure calculates the similarity $s(u,v)$ between a node $u\in \mathcal{G}^{\mathbf{d}}$ and $v\in \mathcal{G}^{\mathbf{d}'}$ as follows. \ding{182} We perform many random walks from node $u$ on graph $\mathcal{G}^{\mathbf{d}}$ and from node $v$ on graph $\mathcal{G}^{\mathbf{d}'}$, and count the number of random walks that stop at each node in $\mathcal{T}$, which can be represented as two \textit{stop count vectors} $\mathbf{c}_u$ and $\mathbf{c}_v$. \ding{183} We calculate the cosine similarity of the count vectors as the similarity between node $u$ and $v$ as follows
\begin{equation}
    \label{eq: node_sim}
    s(u, v) = \frac{\mathbf{c}_u^{\top} \cdot \mathbf{c}_v}{\|\mathbf{c}_u \|_2 \cdot \|\mathbf{c}_v\|_2} = \frac{\mathbf{c}_u^{\top} }{\|\mathbf{c}_u \|_2} \cdot \frac{\mathbf{c}_v}{\|\mathbf{c}_v\|_2} = \hat{\mathbf{c}}_u^{\top}\hat{\mathbf{c}}_v,
\end{equation}
where $\hat{\mathbf{c}}_u$ and $\hat{\mathbf{c}}_v$ are the normalized stop count vectors. When either $\mathbf{c}_u$ or $\mathbf{c}_v$ is a zero vector, $s(u, v)$ will be undefined, and we set $s(u, v) = 0$ in this case.



The rationale behind our cross-domain node similarity measure is that nodes similar to the same set of nodes should be similar. It means the nodes (users or items) may exhibit similar behavior patterns or properties even though they are in different domains. In graph processing, a node $v$ is considered similar (or adjacent) to another node $v'$ if random walks starting from $v$ are likely to reach $v'$~\cite{LiSDCC22, XiongY22}. Thus, the normalized stop count vector $\hat{\mathbf{c}}_v$ measures the similarity between node $v$ and the common nodes in $\mathcal{T}$, and $s(u,v)=\hat{\mathbf{c}_u}^{\top}\hat{\mathbf{c}_v}$ is large if $u$ and $v$ are well-aligned in their similarities between the common nodes.

Armed with the node similarity measure, we find a set $\mathcal{S}_{\mathbf{d}\mathbf{d}'}$ of similar nodes for each pair of domain $\mathbf{d}$ and $\mathbf{d}'$. In particular, for each node $u\in \mathbf{d}$, we compute its similarity between all nodes in $\mathbf{d}'$, and add the $u$ and its top-$k$ similar nodes to $\mathcal{S}_{\mathbf{d}\mathbf{d}'}$ in the form of similar node pairs $(u,v)$. 
As finding similar nodes requires the similarity between all node pairs, its complexity can be high when the domains contain many nodes. However, the task boils down to an inner product-based similarity search for the normalized stop count vectors, which can be conducted efficiently via well-studied vector similarity search solutions~\cite{LSH, HNSW}. Besides, the random walks of each node can also be conducted in parallel.

It is worth noting that we do not directly add all common nodes (overlapping users or items) between domain $\mathbf{d}$ and $\mathbf{d}'$ into the $\mathcal{S}_{\mathbf{d}\mathbf{d}'}$ because the same user/item may also have different behavior patterns or properties in different domains (i.e., intra-domain knowledge). They are not necessarily similar. We also did experiments and found that adding the overlapping users/items to similar node pairs will not improve the performance. 

The similar node pairs found by domain alignment are meaningful. For example, in our experiments on the Amazon dataset, Prism (a Japan CD, in the [Digital Music] domain) is identified to be similar to Shakuhachi (a Japanese musical instrument, in the [Musical Instrument] domain). 

\stitle{Exploit alignment} 
Similar node pairs can be used to assist learning in many possible ways. Currently, we constrain that a similar node pair $(u,v)$ should have similar intra-domain embeddings.
However, since the embeddings belonging to different domains exist in different embedding spaces, directly measuring the distance between them and forcing them to be closer is difficult and not proper. To address this, we employ a linear mapping function~\cite{TransH} to project both embeddings into a shared embedding space. We measure the similarity of the embeddings using the squared Euclidean norm defined as $l(u, v) = \| \boldsymbol{e}^{\mathbf{d}}_{u} W_\mathbf{d} - \boldsymbol{e}^{\mathbf{d}'}_{v} W_{\mathbf{d}'} \|_2^2$, where $W_\mathbf{d}$ and $W_{\mathbf{d}'}$ are the projection matrices for domain $\mathbf{d}$ and $\mathbf{d}'$, respectively. We incorporate the following alignment term in the training objective:

\begin{equation}
    L_{\text{align}} = \sum_{\mathbf{d}, \mathbf{d}' \in \mathbf{D}, \mathbf{d}\neq \mathbf{d}'} \sum_{(u, v) \in \mathcal{S}_{\mathbf{d}\mathbf{d}'}} l(u,v). 
    \label{eq: align}
\end{equation}

Instead of using only the top-$k$ similar node pairs (i.e., $\mathcal{S}_{\mathbf{d}\mathbf{d}'}$) as in Eq.~\eqref{eq: cf}, an alternative is to consider all possible node pairs and weigh $l(u, v)$ by Eq.~\eqref{eq: node_sim}. This design also encourages the embeddings of more similar nodes to better match each other but we do not adopt it because there is a giant number of possible node pairs, which hinders scalability and makes training slow. Combining Eq.~\eqref{eq: cf} and \eqref{eq: align}, our overall loss function is 

\begin{equation}
    L = L_{\text{BPR}} + \beta L_{\text{align}} +\lambda\|\Theta\|_2^2,
\end{equation}

where $\beta$ is a positive weight for the alignment loss, $\Theta$ denotes the trainable parameters in \name, and $\lambda$ is the weight of the $L_2$ regularization. We initialize the intra-domain and inter-domain embeddings individually and train with the Adam optimizer~\cite{AdamKingmaB14}.

\section{Experimental Evaluation} \label{sec:exp}

In this part, we introduce the experiment settings in Section~\ref{subsec:expset}, compare our \name~ model for MDR with state-of-the-art baselines in Section~\ref{subsec:main exp}, and evaluate the key designs of \name~ in Section~\ref{subsec:abl}.

\begin{table}[]
    \caption{Statistics of the datasets, where \textit{overlap} is the ratio of the overlapping users/items between two domains over all users/items in the domains, averaged over all domain pairs.}
    \label{tab:stat}
    \resizebox{0.45\textwidth}{!}{%
        \begin{tabular}{c|c|c|c|c|c}
        \hline
        Dataset & \#Domain   & \#User  & \#Item  & \#Interaction & Overlap  \\ \hline
        AliCCP  & 3          & 47,000  & 53,201  & 353,150       & 0.0190   \\ 
        Amazon  & 6          & 173,528 & 69,215  & 1,437,381     & 0.0830   \\
        AliAd   & 8          & 154,607 & 16,607  & 406,952       & 0.1346   \\ \hline
        \end{tabular}
    }
    \vspace{-1em}
\end{table}

\begin{table*}[]
    \centering
    \caption{Accuracy comparison on three datasets. We mark the best results in bold and the second-best ones in underline. "*" denotes that the best-performing method significantly outperforms the second-best one on the paired t-test ($p$-value < 0.05). }
    \label{tab:per-all}
    \resizebox{0.95\textwidth}{!}{%
    \begin{tabular}{|c|c|c|ccc|ccc|cccccc|c|c|}
    \hline
    \multicolumn{1}{|l|}{}   & \multicolumn{1}{l|}{}   & \multicolumn{1}{l|}{} & MF    & NGCF  & LightGCN & CS    & MMoE  & PLE   & AFT   & STAR  & SAML  & CATART & MGFN        & TreeMS      & \name          & Impr. \\ \hline
    \multirow{8}{*}{AliCCP}  & \multirow{4}{*}{AUC}    & D1                    & 0.562 & 0.661 & 0.720    & 0.591 & 0.639 & 0.629 & 0.560 & 0.597 & 0.572 & 0.547  & 0.762       & {\ul 0.790} & \textbf{0.835}* & 5.6\%   \\
                            &                         & D2                    & 0.555 & 0.595 & 0.590    & 0.639 & 0.582 & 0.624 & 0.547 & 0.570 & 0.567 & 0.501  & {\ul 0.805} & 0.801       & \textbf{0.866}* & 7.6\%   \\
                            &                         & D3                    & 0.648 & 0.721 & 0.766    & 0.582 & 0.608 & 0.603 & 0.557 & 0.581 & 0.584 & 0.593  & 0.768       & {\ul 0.818} & \textbf{0.846}* & 3.4\%   \\
                            &                         & AVG                   & 0.588 & 0.659 & 0.692    & 0.604 & 0.610 & 0.619 & 0.554 & 0.583 & 0.574 & 0.547  & 0.778       & {\ul 0.803} & \textbf{0.849}* & 5.7\%   \\ \cline{2-17} 
                            & \multirow{4}{*}{Recall} & D1                    & 0.147 & 0.241 & 0.332    & 0.168 & 0.204 & 0.190 & 0.114 & 0.171 & 0.156 & 0.147  & 0.179       & {\ul 0.390} & \textbf{0.492}* & 26.3\%  \\
                            &                         & D2                    & 0.204 & 0.184 & 0.132    & 0.204 & 0.182 & 0.184 & 0.082 & 0.122 & 0.122 & 0.102  & {\ul 0.403} & 0.388       & \textbf{0.571}* & 41.8\%  \\
                            &                         & D3                    & 0.251 & 0.303 & 0.380    & 0.155 & 0.233 & 0.209 & 0.106 & 0.153 & 0.152 & 0.185  & 0.176       & {\ul 0.423} & \textbf{0.510}* & 20.7\%  \\
                            &                         & AVG                   & 0.201 & 0.243 & 0.278    & 0.176 & 0.206 & 0.194 & 0.100 & 0.149 & 0.143 & 0.145  & 0.252       & {\ul 0.400} & \textbf{0.524}* & 31.0\%  \\ \hline
    \multirow{14}{*}{Amazon} & \multirow{7}{*}{AUC}    & D1                    & 0.763 & 0.843 & 0.861    & 0.757 & 0.772 & 0.759 & 0.665 & 0.742 & 0.829 & 0.809  & 0.868       & {\ul 0.872} & \textbf{0.904}* & 3.7\%   \\
                            &                         & D2                    & 0.749 & 0.808 & 0.823    & 0.761 & 0.778 & 0.769 & 0.680 & 0.732 & 0.771 & 0.799  & 0.835       & {\ul 0.835} & \textbf{0.871}* & 4.3\%   \\
                            &                         & D3                    & 0.764 & 0.826 & 0.855    & 0.785 & 0.710 & 0.789 & 0.583 & 0.740 & 0.713 & 0.747  & 0.868       & {\ul 0.869} & \textbf{0.893}* & 2.8\%   \\
                            &                         & D4                    & 0.721 & 0.731 & 0.749    & 0.754 & 0.766 & 0.760 & 0.649 & 0.705 & 0.730 & 0.719  & 0.761       & {\ul 0.768} & \textbf{0.796}* & 3.6\%   \\
                            &                         & D5                    & 0.784 & 0.857 & 0.882    & 0.771 & 0.788 & 0.768 & 0.677 & 0.746 & 0.759 & 0.761  & 0.894       & {\ul 0.897} & \textbf{0.919}* & 2.4\%   \\
                            &                         & D6                    & 0.842 & 0.862 & 0.876    & 0.823 & 0.846 & 0.844 & 0.640 & 0.668 & 0.757 & 0.781  & 0.879       & {\ul 0.889} & \textbf{0.905}* & 1.8\%   \\
                            &                         & AVG                   & 0.771 & 0.821 & 0.841    & 0.775 & 0.777 & 0.782 & 0.649 & 0.722 & 0.760 & 0.770  & 0.851       & {\ul 0.855} & \textbf{0.881}* & 3.1\%   \\ \cline{2-17} 
                            & \multirow{7}{*}{Recall} & D1                    & 0.456 & 0.568 & 0.595    & 0.513 & 0.511 & 0.514 & 0.262 & 0.497 & 0.472 & 0.504  & 0.611       & {\ul 0.618} & \textbf{0.677}* & 9.6\%   \\
                            &                         & D2                    & 0.422 & 0.493 & 0.550    & 0.432 & 0.433 & 0.437 & 0.287 & 0.444 & 0.532 & 0.479  & 0.568       & {\ul 0.573} & \textbf{0.601}* & 4.9\%   \\
                            &                         & D3                    & 0.473 & 0.521 & 0.591    & 0.396 & 0.403 & 0.404 & 0.151 & 0.369 & 0.415 & 0.433  & 0.605       & {\ul 0.613} & \textbf{0.641}* & 4.6\%   \\
                            &                         & D4                    & 0.327  & 0.365 & 0.404    & 0.363 & 0.369 & 0.365 & 0.221 & 0.346 & 0.354 & 0.355  & 0.419       & {\ul 0.421} & \textbf{0.457}* & 8.4\%   \\
                            &                         & D5                    & 0.441 & 0.578 & 0.596    & 0.502 & 0.499 & 0.502 & 0.265 & 0.440 & 0.578 & 0.550  & 0.610       & {\ul 0.629} & \textbf{0.699}* & 11.1\%  \\
                            &                         & D6                    & 0.533 & 0.534 & 0.572    & 0.533 & 0.550 & 0.595 & 0.236 & 0.434 & 0.588 & 0.600  & 0.603       & {\ul 0.606} & \textbf{0.618}* & 2.1\%   \\
                            &                         & AVG                   & 0.444 & 0.510 & 0.551    & 0.457 & 0.461 & 0.470 & 0.237 & 0.422 & 0.490 & 0.487  & 0.569       & {\ul 0.577} & \textbf{0.616}* & 6.7\%   \\ \hline
    \multirow{18}{*}{AliAd}  & \multirow{9}{*}{AUC}    & D1                    & 0.803 & 0.832 & 0.843    & 0.820 & 0.833 & 0.841 & 0.725 & 0.844 & 0.808 & 0.809  & 0.852       & {\ul 0.856} & \textbf{0.877}* & 2.4\%   \\
                            &                         & D2                    & 0.873 & 0.885 & 0.896    & 0.888 & 0.885 & 0.900 & 0.817 & 0.882 & 0.888 & 0.876  & 0.909       & {\ul 0.910} & \textbf{0.931}* & 2.3\%   \\
                            &                         & D3                    & 0.742 & 0.746 & 0.750    & 0.785 & 0.814 & 0.809 & 0.707 & 0.808 & 0.753 & 0.745  & 0.780       & {\ul 0.818} & \textbf{0.854}* & 4.4\%   \\
                            &                         & D4                    & 0.899 & 0.895 & 0.895    & 0.886 & 0.883 & 0.903 & 0.825 & 0.889 & 0.889 & 0.895  & 0.900       & {\ul 0.911} & \textbf{0.922}* & 1.2\%   \\
                            &                         & D5                    & 0.810 & 0.835 & 0.837    & 0.836 & 0.838 & 0.852 & 0.763 & 0.850 & 0.798 & 0.791  & 0.850       & {\ul 0.863} & \textbf{0.913}* & 5.8\%   \\
                            &                         & D6                    & 0.824 & 0.826 & 0.830    & 0.847 & 0.852 & 0.860 & 0.779 & 0.854 & 0.782 & 0.786  & 0.846       & {\ul 0.861} & \textbf{0.897}* & 4.2\%   \\
                            &                         & D7                    & 0.815 & 0.822 & 0.835    & 0.834 & 0.836 & 0.843 & 0.756 & 0.845 & 0.789 & 0.781  & 0.844       & {\ul 0.852} & \textbf{0.903}* & 6.0\%   \\
                            &                         & D8                    & 0.833 & 0.863 & 0.868    & 0.868 & 0.861 & 0.874 & 0.766 & 0.864 & 0.842 & 0.834  & 0.866       & {\ul 0.889} & \textbf{0.921}* & 3.6\%   \\
                            &                         & AVG                   & 0.825 & 0.837 & 0.844    & 0.846 & 0.850 & 0.860 & 0.767 & 0.855 & 0.819 & 0.814  & 0.856       & {\ul 0.870} & \textbf{0.902}* & 3.7\%   \\ \cline{2-17} 
                            & \multirow{9}{*}{Recall} & D1                    & 0.437 & 0.499 & 0.565    & 0.533 & 0.537 & 0.548 & 0.281 & 0.541 & 0.575 & 0.448  & 0.566       & {\ul 0.579} & \textbf{0.617}* & 6.6\%   \\
                            &                         & D2                    & 0.574 & 0.593 & 0.655    & 0.579 & 0.554 & 0.576 & 0.449 & 0.571 & 0.673 & 0.585  & 0.672       & {\ul 0.678} & \textbf{0.728}* & 7.4\%   \\
                            &                         & D3                    & 0.382 & 0.379 & 0.472    & 0.389 & 0.428 & 0.410 & 0.261 & 0.392 & 0.454 & 0.374  & 0.489       & {\ul 0.493} & \textbf{0.593}* & 20.3\%  \\
                            &                         & D4                    & 0.587 & 0.618 & 0.628    & 0.561 & 0.555 & 0.588 & 0.458 & 0.579 & 0.629 & 0.593  & 0.632       & {\ul 0.642} & \textbf{0.681}* & 6.1\%   \\
                            &                         & D5                    & 0.467 & 0.529 & 0.583    & 0.456 & 0.465 & 0.466 & 0.364 & 0.464 & 0.551 & 0.475  & 0.599       & {\ul 0.604} & \textbf{0.687}* & 13.8\%  \\
                            &                         & D6                    & 0.509 & 0.513 & 0.599    & 0.513 & 0.510 & 0.511 & 0.383 & 0.493 & 0.591 & 0.475  & 0.611       & {\ul 0.627} & \textbf{0.680}* & 8.5\%   \\
                            &                         & D7                    & 0.482 & 0.480 & 0.604    & 0.460 & 0.462 & 0.463 & 0.340 & 0.454 & 0.574 & 0.469  & 0.615       & {\ul 0.619} & \textbf{0.683}* & 10.2\%  \\
                            &                         & D8                    & 0.501 & 0.550 & 0.612    & 0.500 & 0.494 & 0.509 & 0.371 & 0.499 & 0.539 & 0.562  & 0.639       & {\ul 0.653} & \textbf{0.690}* & 5.7\%   \\
                            &                         & AVG                   & 0.492 & 0.520 & 0.590    & 0.499 & 0.501 & 0.509 & 0.363 & 0.499 & 0.573 & 0.497  & 0.603       & {\ul 0.612} & \textbf{0.670}* & 9.5\%   \\ \hline
    \end{tabular}%
    }
    \vspace{-0.5em}
    \end{table*}

\subsection{Experiment Settings}\label{subsec:expset}

\stitle{Datasets}
We adopt three commonly-used public datasets to evaluate MDR solutions, and their statistics are reported in Table~\ref{tab:stat}.
\begin{itemize}
    \item \textbf{AliCCP}~\cite{ESMM} gathers traffic logs of the recommender system in mobile Taobao, which contains three different domains.
    \item \textbf{Amazon}~\cite{Amazon} contains user ratings for items from Amazon. We use 6 categories of items 
    and regard each category as a domain. Each rating is considered a positive interaction. 
    \item \textbf{AliAd}~\cite{AliAd} consists of 8 days of ad display/click records from Alibaba. We choose 8 categories of ads as different domains and regard clicks as positive interactions.
\end{itemize}

%

\stitle{Baselines} 
For a comprehensive comparison, we compare our \name~ with 12 state-of-the-art baseline methods including 3 single-domain recommendation (SDR) models (i.e., \textbf{MF}~\cite{abs-1205-2618}, \textbf{NGCF}~\cite{NGCF}, \textbf{LightGCN}~\cite{0001DWLZ020}), 3 multi-task learning (MTL) methods (i.e.,  \textbf{Cross-Stitch (CS)}~\cite{MisraSGH16}, \textbf{MMoE}~\cite{MMoEMaZYCHC18}, \textbf{PLE}~\cite{PLETangLZG20}), and 6 MDR methods (i.e., \textbf{AFT}~\cite{AFTHaoLXGTZL21}, \textbf{STAR}~\cite{StarShengZZDDLYLZDZ21}, \textbf{SAML}~\cite{saml}, \textbf{CATART}~\cite{catart}, \textbf{MGFN}~\cite{MGFNZhangPWPZLQ22}, and \textbf{TreeMS}~\cite{TreeMSNiuLLT0D21}). Due to the space limits, we will discuss the MTL and MDR baselines in Section~\ref{sec:rel}. For SDR methods, we train a separate model for each domain. For MTL methods, they handle different tasks on the same data and are usually used in MDR~\cite{StarShengZZDDLYLZDZ21, AFTHaoLXGTZL21, TreeMSNiuLLT0D21} by treating the aggregated data from all domains as the input data and recommendation for each domain as a task. 

\stitle{Performance metrics} 
We focus on the accuracy of recommendation and adopt two widely utilized metrics, i.e., \textbf{AUC}~\cite{StarShengZZDDLYLZDZ21} and \textbf{Recall}~\cite{AFTHaoLXGTZL21}, to evaluate the models. Following previous works~\cite{KGATWang00LC19, LiuCZGN21}, we regard the interacted items as positive for each user and split the interaction records into training/validation/testing sets with a ratio of 7:1:2. For each positive item, we randomly sample 10 un-interacted items as negative items in the test set. Recall is calculated as the percentage of times that the positive item ranks top-1 among the 11 items (i.e., Recall@1)~\cite{RMS}.

\stitle{Implementation details}  
We implement all models using Pytorch, DGL~\cite{DGL} and LibMTL~\cite{LibMTL}. For \disgcn, the default dimension of the inter-domain and intra-domain embeddings are both 64, the number of layers is 2 and $\alpha$ is 0.1 in GRec, the batch size is 8092. The weight of domain alignment loss (i.e., $\beta$) is 0.03, the regularization coefficient $\lambda$ is 1e-4 and learning rate is 0.001. We adopt the edge dropout~\cite{DropEdgeRongHXH20} to train our \name~ and set the dropout ratio as 0.3. For the domain alignment in \name, we set the random walk length as 4 and the number of walks from each node is 500. We select the node with the largest similarity for each node (i.e., $k = 1$) as its similar node by default. We conduct mode training via Adam~\cite{AdamKingmaB14} optimizer. 
For all the baseline methods, we run them using the default settings in their released codes and original reports.


\subsection{Main Results: Compare with the Baselines}\label{subsec:main exp}

Table~\ref{tab:per-all} compares the recommendation accuracy of our \name~ with the 12 baselines on the 3 experiment datasets. \textit{Impr.} is the percentage of accuracy improvement of \name~ over the best-performing baseline, and \textit{Avg} is the average statistics over all domains in a dataset. We make the following observations from the results.

Firstly, our \name~ consistently outperforms all baselines across all datasets and domains, achieving higher AUC and recall values. The improvements of \name~ over the best-performing baseline are significant, reaching up to 7.6\% and 41.8\% for AUC and recall, respectively, in specific domains. Overall, \name~ exhibits more significant improvements over the baselines for AliCCP (5.7\% in AUC and 31.0\% in recall for average) compared to Amazon (3.1\% in AUC and 6.7\% in recall for average) and AliAd (3.7\% in AUC and 9.5\% in recall for average). This can be attributed to the smaller size of the AliCCP dataset (i.e., fewer interaction records) and the lower overlap between domains (as shown in Table~\ref{tab:stat}). Consequently, AliCCP benefits more from \name's ability to complement the domains, while the larger domain overlap in the Amazon and AliAd datasets already enables more knowledge sharing among domains, making the benefits of \name~ comparatively less significant.

\begin{table*}[]
    \centering
    \caption{Ablation study on the three datasets. We mark the best results in bold and the second-best ones in underline.}
    \label{tab:abl-all}
    \resizebox{0.8\textwidth}{!}{%
    
\begin{tabular}{|c|c|cccccccccccc|}
\hline
\multirow{10}{*}{AUC}    & \textbf{AliCCP} & D1             & D2             & D3             & \multicolumn{1}{c|}{AVG}            & \multicolumn{1}{c|}{\textbf{Amazon}} & D1             & D2             & D3             & D4             & D5             & D6             & AVG            \\ \cline{2-14} 
                         & Inter           & 0.815          & {\ul 0.865}    & 0.829          & \multicolumn{1}{c|}{0.836}          & \multicolumn{1}{c|}{Inter}           & 0.869          & 0.828          & 0.850          & 0.733          & 0.878          & 0.879          & 0.840          \\
                         & Intra           & 0.769          & 0.567          & 0.824          & \multicolumn{1}{c|}{0.720}          & \multicolumn{1}{c|}{Intra}           & 0.878          & 0.844          & 0.869          & 0.760          & 0.889          & 0.865          & 0.851          \\
                         & w/o DA          & {\ul 0.822}    & 0.864          & {\ul 0.835}    & \multicolumn{1}{c|}{{\ul 0.840}}    & \multicolumn{1}{c|}{w/o DA}          & {\ul 0.886}    & {\ul 0.856}    & {\ul 0.881}    & {\ul 0.775}    & {\ul 0.905}    & {\ul 0.891}    & {\ul 0.866}    \\
                         & \name           & \textbf{0.835} & \textbf{0.866} & \textbf{0.846} & \multicolumn{1}{c|}{\textbf{0.849}} & \multicolumn{1}{c|}{\name}           & \textbf{0.904} & \textbf{0.871} & \textbf{0.893} & \textbf{0.796} & \textbf{0.919} & \textbf{0.905} & \textbf{0.881} \\ \cline{2-14} 
                         & \textbf{AliAd}  & D1             & D2             & D3             & D4                                  & D5                                   & D6             & D7             & D8             & AVG            &                &                &                \\ \cline{2-14} 
                         & Inter           & 0.822          & 0.858          & 0.793          & 0.850                               & 0.870                                & 0.848          & 0.857          & 0.857          & 0.844          &                &                &                \\
                         & Intra           & 0.846          & 0.891          & 0.806          & 0.887                               & 0.877                                & 0.851          & 0.857          & 0.885          & 0.863          &                &                &                \\
                         & w/o DA          & {\ul 0.867}    & {\ul 0.919}    & {\ul 0.828}    & {\ul 0.908}                         & {\ul 0.899}                          & {\ul 0.878}    & {\ul 0.882}    & {\ul 0.902}    & {\ul 0.885}    &                &                &                \\
                         & \name           & \textbf{0.877} & \textbf{0.931} & \textbf{0.854} & \textbf{0.922}                      & \textbf{0.913}                       & \textbf{0.897} & \textbf{0.903} & \textbf{0.921} & \textbf{0.902} &                &                &                \\ \hline
\multirow{10}{*}{Recall} & \textbf{AliCCP} & D1             & D2             & D3             & \multicolumn{1}{c|}{AVG}            & \multicolumn{1}{c|}{\textbf{Amazon}} & D1             & D2             & D3             & D4             & D5             & D6             & AVG            \\ \cline{2-14} 
                         & Inter           & 0.229          & 0.509          & 0.226          & \multicolumn{1}{c|}{0.321}          & \multicolumn{1}{c|}{Inter}           & 0.262          & 0.252          & 0.228          & 0.187          & 0.258          & 0.265          & 0.242          \\
                         & Intra           & 0.391          & 0.184          & 0.472          & \multicolumn{1}{c|}{0.349}          & \multicolumn{1}{c|}{Intra}           & 0.651          & 0.574          & 0.612          & 0.434          & 0.675          & 0.570          & 0.586          \\
                         & w/o DA          & {\ul 0.481}    & {\ul 0.551}    & {\ul 0.490}    & \multicolumn{1}{c|}{{\ul 0.507}}    & \multicolumn{1}{c|}{w/o DA}          & {\ul 0.662}    & {\ul 0.589}    & {\ul 0.626}    & {\ul 0.446}    & {\ul 0.682}    & {\ul 0.589}    & {\ul 0.599}    \\
                         & \name           & \textbf{0.492} & \textbf{0.571} & \textbf{0.509} & \multicolumn{1}{c|}{\textbf{0.524}} & \multicolumn{1}{c|}{\name}           & \textbf{0.677} & \textbf{0.601} & \textbf{0.641} & \textbf{0.457} & \textbf{0.699} & \textbf{0.624} & \textbf{0.616} \\ \cline{2-14} 
                         & \textbf{AliAd}  & D1             & D2             & D3             & D4                                  & D5                                   & D6             & D7             & D8             & AVG            &                &                &                \\ \cline{2-14} 
                         & Inter           & 0.305          & 0.418          & 0.385          & 0.279                               & 0.452                                & 0.438          & 0.437          & 0.384          & 0.387          &                &                &                \\
                         & Intra           & 0.588          & 0.673          & 0.526          & 0.647                               & 0.650                                & 0.626          & 0.632          & 0.642          & 0.623          &                &                & {\ul }         \\
                         & w/o DA          & {\ul 0.600}    & {\ul 0.702}    & {\ul 0.557}    & {\ul 0.670}                         & {\ul 0.673}                          & {\ul 0.652}    & {\ul 0.661}    & {\ul 0.662}    & {\ul 0.647}    &                &                &                \\
                         & \name           & \textbf{0.617} & \textbf{0.728} & \textbf{0.593} & \textbf{0.681}                      & \textbf{0.687}                       & \textbf{0.680} & \textbf{0.683} & \textbf{0.690} & \textbf{0.670} &                &                &                \\ \hline
\end{tabular}
    }
    \vspace{-1em}
    \end{table*}

Secondly, compared with the best-performing baseline, the accuracy improvements of \name~ are similar for different domains on Amazon but differ significantly for AliCCP (with AliAd somewhere between the two extremes). In particular, on Amazon, the recall improvements are around 2\%--11\% for all domains; For AliCCP, recall improvements are  41.8\% and 20.7\% for D2 and D3, respectively. This is because AliCCP has more skewed domain sizes, and \name~ generally yields larger gains for smaller domains. In the case of Amazon, the largest domain's size is 16.5 times that of the smallest domain. However, for AliCCP, the size of D3 is 75.7 times that of D2. Smaller domains benefit more from \name~ because they are more susceptible to over-fitting and poor generalization due to their limited training data. As a result, the cross-domain knowledge transfer capability of \name~ becomes particularly crucial for improving the performance of smaller domains.



Thirdly, the graph neural network (GNN)-based baselines (e.g., MGFN, TreeMS) generally achieve higher accuracy better than the MLP-based baselines (e.g., STAR, CS, MMoE, PLE). This is because GNN can capture the high-order dependency between users and items with neighbor aggregation and better capture collaborative filtering signals~\cite{0001DWLZ020, LiuCZGN21, NGCF}. It is also the reason why we use graph-based learning in our model design.  



\subsection{Ablation Study and Insights}\label{subsec:abl}


\stitle{Ablation study} 
We design 3 variants and compare them with \name~ to evaluate our key designs.

\begin{itemize}
    \item \textbf{Inter} only remains the inter-domain model and inter-domain embeddings for recommendation, removing the intra-domain models and embeddings and domain alignment strategy.
    \item \textbf{Intra} trains a separate intra-domain model for each domain using that domain's dataset, removing the inter-domain models and embeddings and domain alignment strategy. 
    \item \textbf{w/o DA} removes the domain alignment loss term (i.e., Eq.~\eqref{eq: align}) from \name~ and only uses ED recommender.
\end{itemize}

For a fair comparison, we set the embedding size in \textit{Inter} and \textit{Intra} as \textbf{twice that in \name}~ to make sure the performance improvements of our ED architecture are not from the increased model parameter quantity. 
Here, comparing \textit{w/o DA} with \textit{Inter} and \textit{Intra} shows the benefits of ED architecture. Comparing \name~ with \textit{w/o DA} helps to understand the effectiveness of our domain alignment strategy. 
Table~\ref{tab:abl-all} report the results of the ablation study, and we make the following observations.

(1) \textit{w/o DA} (which uses ED architecture) outperforms both \textit{Inter} and \textit{Intra} (which do not use ED architecture and consist of only one type of embedding) in most cases, which suggests that ED architecture is effective in improving accuracy. 
(2) The accuracy of \textit{Inter} is usually lower than \textit{Intra}. This indicates that only using a single shared model for all domains is not a good way for cross-domain knowledge sharing, possibly because that specific knowledge for a domain will cause negative transfer in other domains~\cite{WangZJLBMZZ21, LiuLG19}. However, \textit{Inter} also outperforms \textit{Intra} (without cross-domain knowledge sharing) in some cases (e.g., AliCCP D2), which suggests that cross-domain knowledge sharing is beneficial. The good performance of \name~ indicates that ED architecture achieves knowledge sharing and avoids negative transfer at the same time. 
(3) Compared with \textit{w/o DA}, \name~ consistently yields higher accuracy in all cases, highlighting the importance of our domain alignment strategy.

\begin{figure*} 
	\subfloat[AliCCP]{
		\label{fig: AliCCP-case}
		\includegraphics[width=0.3\linewidth]{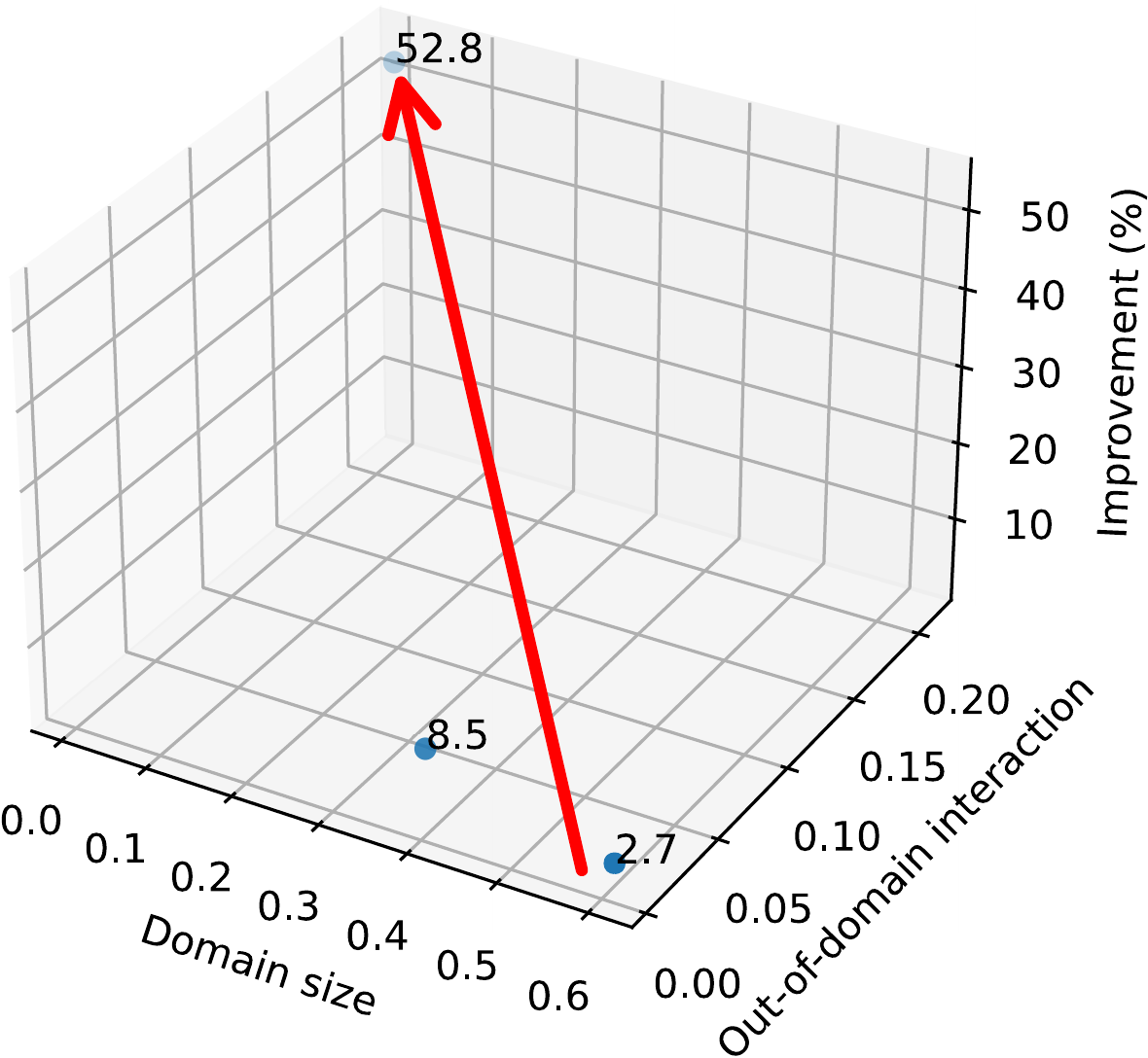}}
	\subfloat[Amazon]{
		\label{fig: amazon-case} 
		\includegraphics[width=0.3\linewidth]{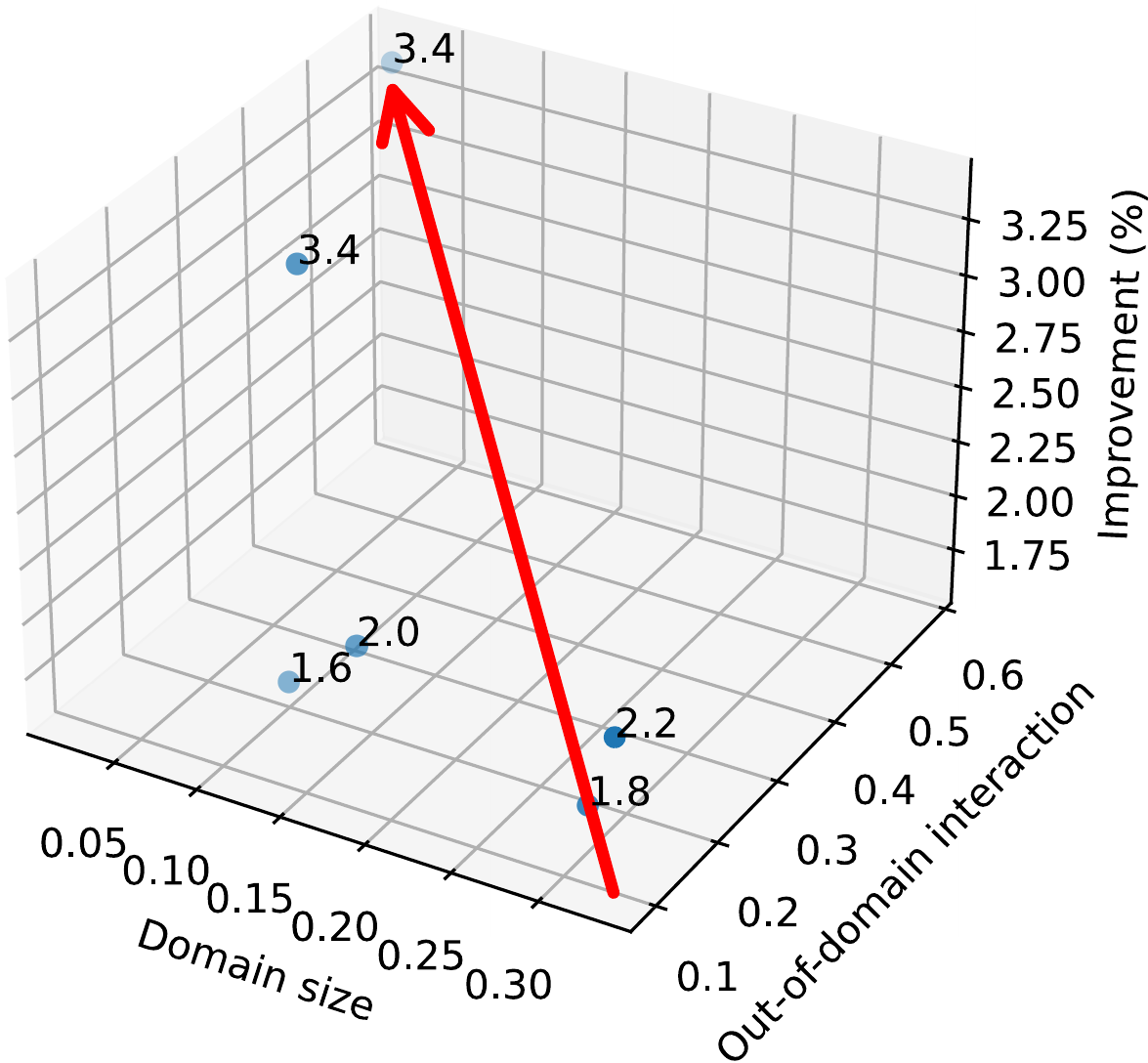}}
	\subfloat[AliAd]{
		\label{fig: AliAd-case}
		\includegraphics[width=0.3\linewidth]{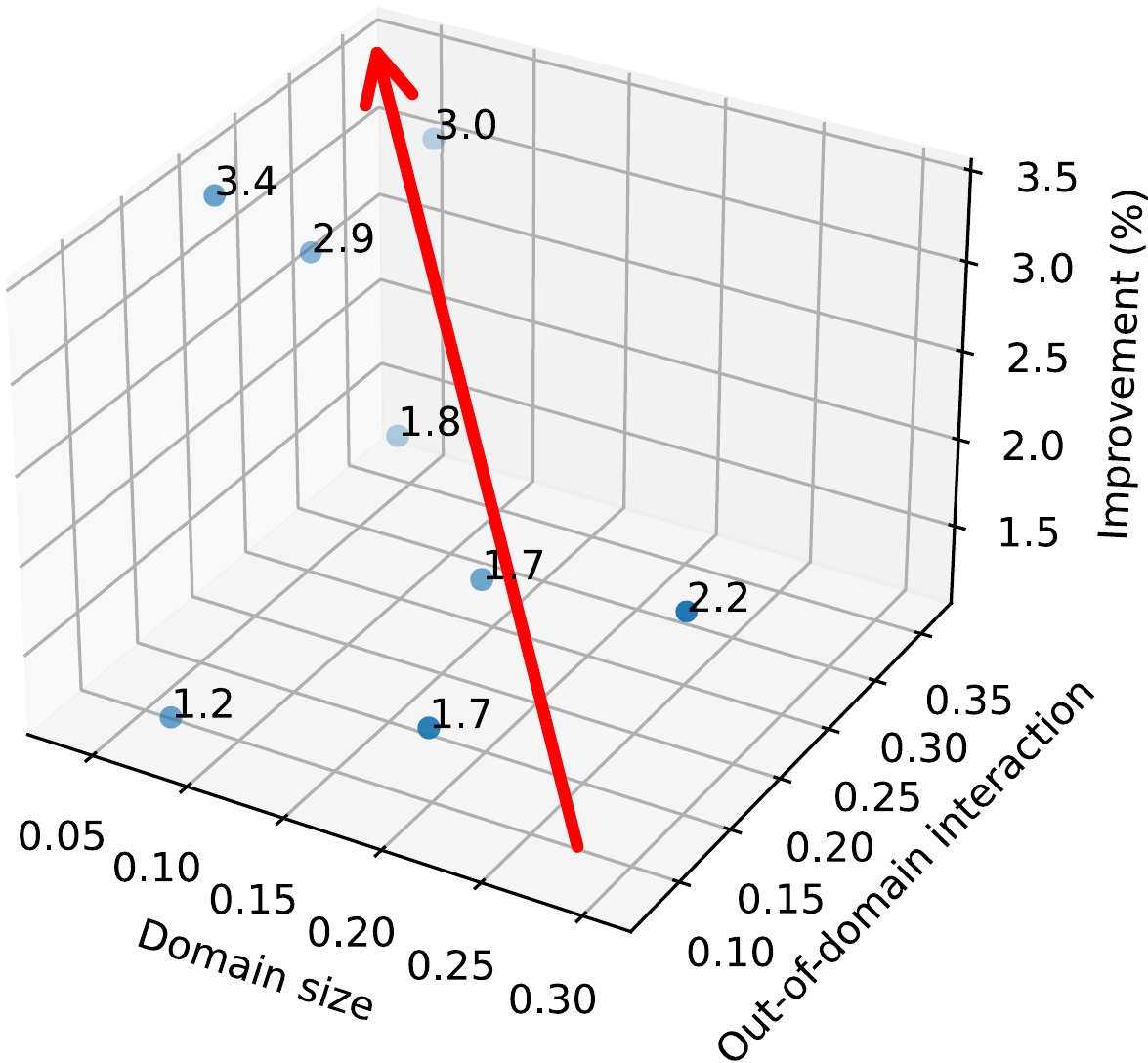}}
	\caption{Relative AUC improvement of \name~ over Intra, each point corresponds to a domain.}
	\label{fig: case} 
	\vspace{-1em}
\end{figure*}

\begin{figure} 
    \subfloat[Amazon]{
    \label{fig: amazon-dis} 
    \includegraphics[width=0.48\linewidth]{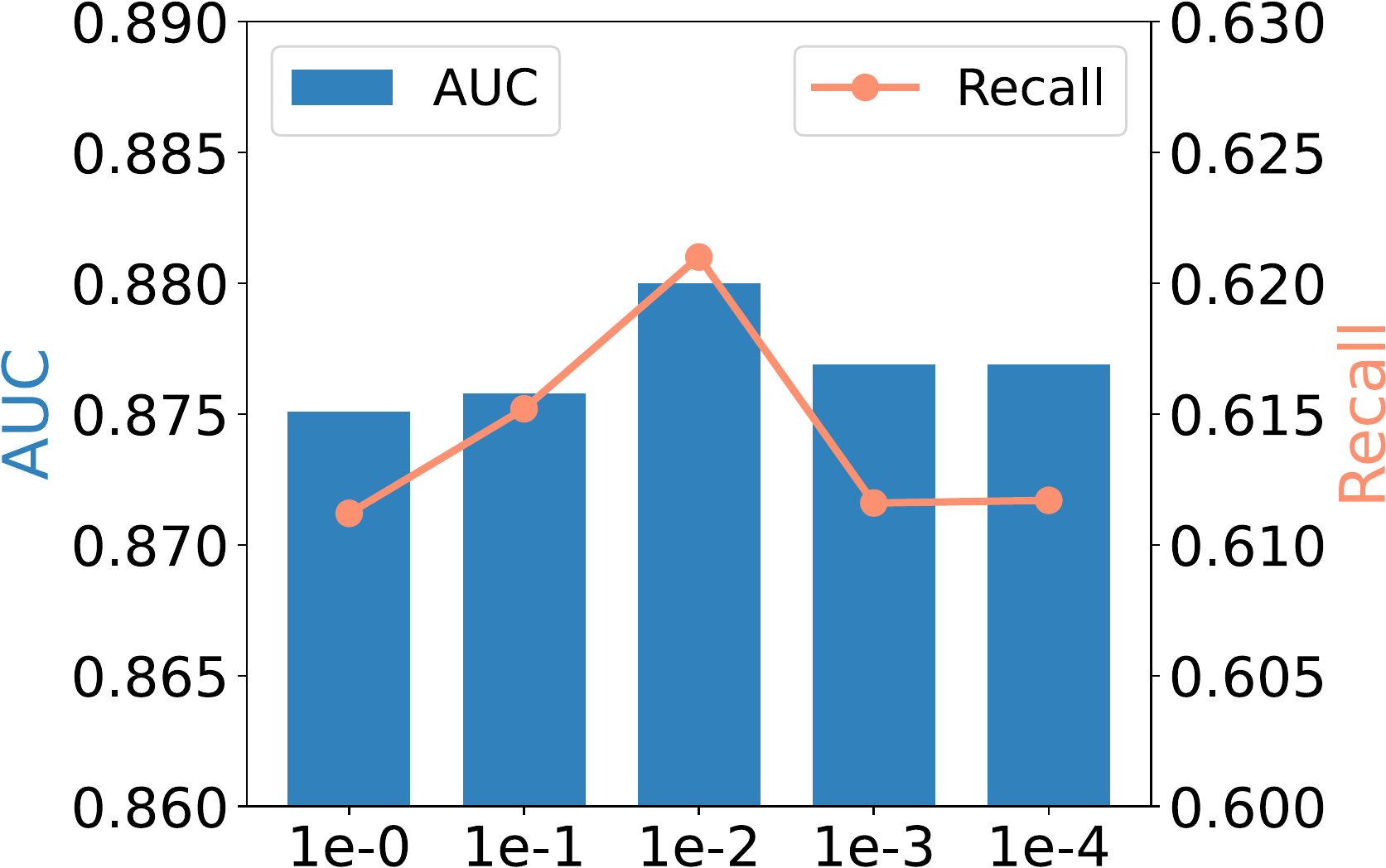}}
    \subfloat[AliAd]{
    \label{fig: AliAd-dis}
    \includegraphics[width=0.48\linewidth]{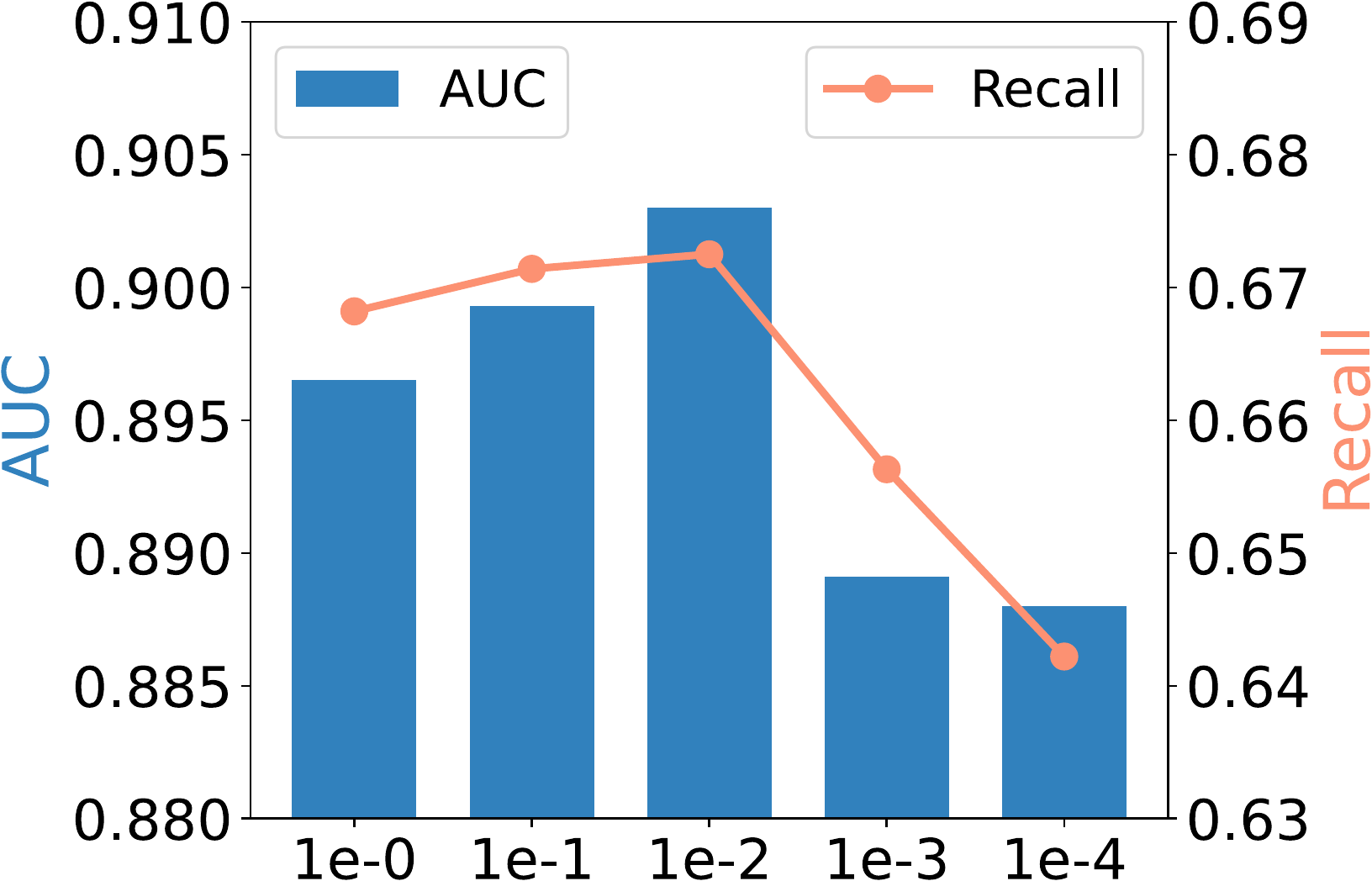}}
    \caption{Effect of weight $\beta$ for alignment loss}
    \label{fig: dis_eff} 
    \vspace{-1.5em}
\end{figure}

\begin{figure}[t]
    \subfloat[Amazon]{
    \label{fig: amazon-emb} 
    \includegraphics[width=0.48\linewidth]{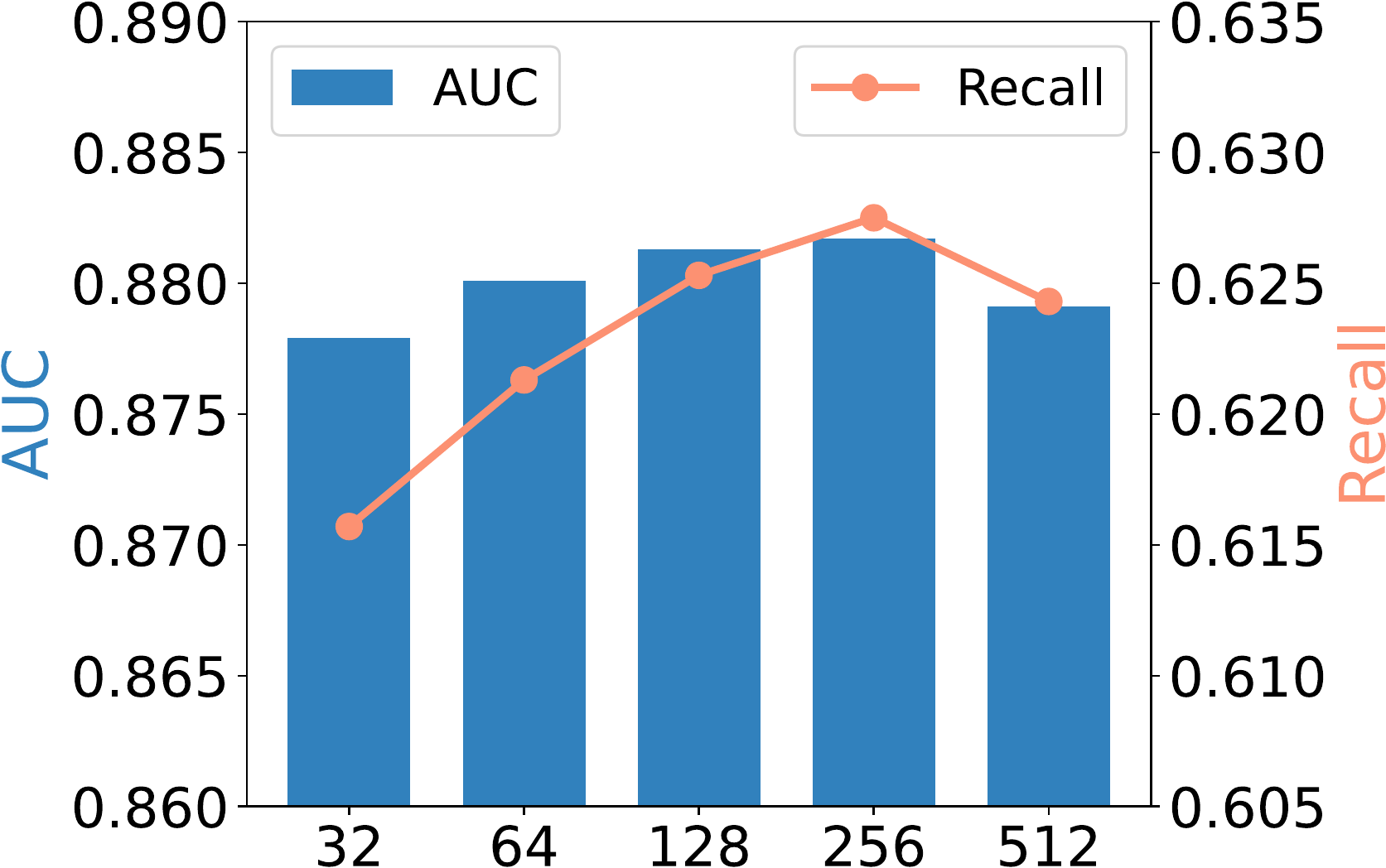}}
    \subfloat[AliAd]{
    \label{fig: AliAd-emb}
    \includegraphics[width=0.48\linewidth]{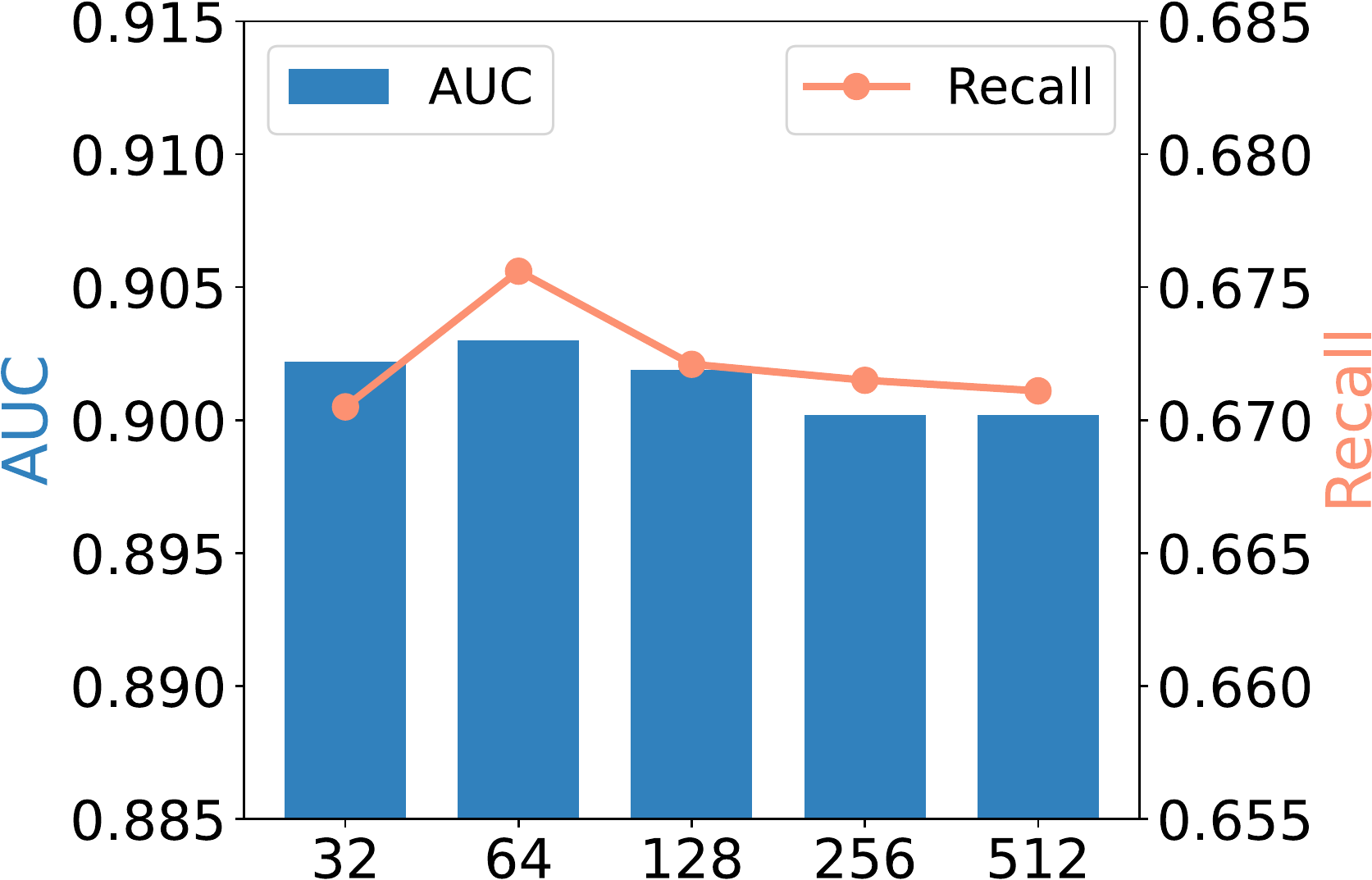}}
    \caption{Effect of embedding dimension}
    \label{fig: emb_eff} 
    \vspace{-1.5em}
\end{figure}

\stitle{Analyzing the gains of \name} 
As a method for MDR, \name~ complements the domains with each other while most recommendation methods handle a single domain. An interesting question is---what are the characteristics of the domains/datasets that benefit the most from \name? We analyze the gains of \name~ over \textit{Intra} (which trains a separate model for each domain) and find that the gains are related to the following two factors for a domain $\mathbf{d}_w$. 
\begin{equation*}
    \begin{aligned}
    	 \text{Domain size:\ } {DS_{\mathbf{d}_w}} &= \frac{|\mathcal{R}^{\mathbf{d}_w}|}{\sum_{\mathbf{d} \in \mathcal{D}} |\mathcal{R}^{\mathbf{d}}|},\\
        \text{Out-of-domain interaction:\ } {OI_{\mathbf{d}_w}} &= \frac{\sum_{u \in \mathcal{U}^{\mathbf{d}_w}} \sum_{\mathbf{d} \in \mathbf{D}, \mathbf{d} \neq \mathbf{d}_w}  |\mathcal{R}_{u}^{\mathbf{d}}|}{|\mathcal{R}^{\mathbf{d}_w}|},
    \end{aligned}
 \end{equation*}
where $\mathcal{R}^{\mathbf{d}_w}$ is the user-item interactions of domain $\mathbf{d}_w$, $\mathcal{R}_{u}^{\mathbf{d}}$ is the interactions made by user $u$ in domain $\mathbf{d}$, and $||$ indicates the cardinality of a set. Intuitively, $DS_{\mathbf{d}_w}$ measures the size of a domain over all domains while $OI_{\mathbf{d}_w}$ measures the interactions of made by users of $\mathbf{d}_w$ in other domains (with a normalization).

Figure~\ref{fig: case} plots the gain of \name~ against the domain size and out-of-domain interaction. For different domains of the same dataset, the gain of \name~ generally increases along the direction of the red arrow, which points to smaller domain sizes and larger out-of-domain interactions. 
Smaller domain sizes indicate that a particular domain has limited training data available. In such cases, the cross-domain knowledge-sharing capability of \name~ becomes especially valuable. By leveraging knowledge from other domains, \name~ can compensate for the lack of data in the small domain and provide more accurate recommendations. 
Larger out-of-domain interaction means that the users make more interactions in other domains, which allows for more knowledge to be transferred. Consequently, \name~ can leverage transferred knowledge to enhance the accuracy.

%




\stitle{Hyper-parameters} 
We adjust the weight $\beta$ for the alignment loss and report the accuracy of \name~ in Figure~\ref{fig: dis_eff}. The results show that the optimal $\beta$ is $1e-2$ for both datasets and accuracy degrades when $\beta$ deviates from the optimal. This is because the alignment loss serves as a regularization that encourages similar user/item pairs from different domains to have similar intra-domain embeddings. When $\beta$ is too small, the strength of the regularization is not enough and information encoded by similar user/item pairs is not fully utilized. When $\beta$ is too large, the regularization affects other learning objectives, e.g., fitting the observed user-item interactions. We adjust the dimension of the intra-domain and inter-domain embeddings and report the accuracy of \name~ in Figure~\ref{fig: dis_eff}. The results show that there exists an optimal embedding dimension for both datasets. This is because the model may not have enough capacity when the dimension is smaller than optimal while over-fitting may happen when the dimension is larger than optimal.

\begin{table*}[]
    \caption{Applying the ED architecture on MF model. 'Impr.' denotes the percentage of improvement.}
    \label{tab:abl-mf}
    \resizebox{0.82\textwidth}{!}{%
    \begin{tabular}{|c|cccccccccccc|}
    \hline
    \textbf{AliCCP} & D1      & D2     & D3     & \multicolumn{1}{c|}{AVG}    & \multicolumn{1}{c|}{\textbf{Amazon}} & D1     & D2     & D3     & D4     & D5                   & D6                   & AVG                   \\ \hline
    MF              & 0.147   & 0.204  & 0.251  & \multicolumn{1}{c|}{0.201}  & \multicolumn{1}{c|}{MF}              & 0.456  & 0.422  & 0.473  & 0.327  & 0.441                & 0.533                & 0.442                 \\
    ED\_MF          & 0.302   & 0.225  & 0.350  & \multicolumn{1}{c|}{0.292}  & \multicolumn{1}{c|}{ED\_MF}          & 0.599  & 0.478  & 0.571  & 0.328  & 0.604                & 0.574                & 0.526                 \\
    Impr.         & 105.2\% & 10.2\% & 39.4\% & \multicolumn{1}{c|}{45.6\%} & \multicolumn{1}{c|}{Impr.}         & 31.3\% & 13.2\% & 20.8\% & 0.3\%  & 36.9\%               & 7.8\%                & 18.9\%                \\ \hline
    \textbf{AliAd}  & D1      & D2     & D3     & D4                          & D5                                   & D6     & D7     & D8     & AVG    &                      &                      &                       \\ \hline
    MF              & 0.437   & 0.574  & 0.382  & 0.587                       & 0.467                                & 0.509  & 0.482  & 0.501  & 0.492  &                      &                      &                       \\
    ED\_MF          & 0.502   & 0.634  & 0.455  & 0.589                       & 0.556                                & 0.550  & 0.545  & 0.571  & 0.550  &                      &                      &                       \\
    Impr.         & 15.0\%  & 10.4\% & 19.3\% & 0.4\%                       & 19.0\%                               & 8.0\%  & 13.0\% & 14.0\% & 11.8\% & \multicolumn{1}{l}{} & \multicolumn{1}{l}{} & \multicolumn{1}{l|}{} \\ \hline
    \end{tabular}
    }
    \vspace{-0.5em}
\end{table*}

\stitle{Other instantiation of ED architecture} 
As discussed, our ED architecture is general and can be applied to any model that learns user and item embeddings. Besides our proposed intra-domain and inter-domain model GRec, we also apply the classical matrix factorization (MF) model~\cite{abs-1205-2618} in ED architecture (denoted as ED\_MF) and compare it with learning a separate MF model for each domain (which performs much better than learning a single MF model using the aggregated data from all domains). We only report recall in Table~\ref{tab:abl-mf} due to page limit. The results show that ED architecture consistently improves the accuracy of MF on all datasets and their domains, and the improvements are usually large, i.e., over 10\% for most cases. Compared to the improvement of \name~ over \textit{Intra} model shown in our ablation study (see Table~\ref{tab:abl-all}), ED architecture yields more significant improvement for MF because MF is a weaker model and thus there is more room for improvement. 




\section{Related Work} \label{sec:rel}

\stitle{Multi-domain recommendation}
Multi-domain recommendation (MDR)~\cite{MAMDR, NNM, DPA, CEMDR} is a category of cross-domain recommendation that focuses on improving the accuracy of multiple domains simultaneously. For example,
AFT~\cite{AFTHaoLXGTZL21} uses user-item interaction data from all domains to learn user interests and adopts a generator-discriminator framework.
These methods leverage the model disentangling architecture and cannot separate domain-shared and domain-specific knowledge well. 
CATART~\cite{catart} uses an auto-encoder to construct user global embeddings from each domain's domain-specific embedding and uses an attention mechanism to aggregate global embeddings and domain-specific embeddings to conduct recommendations for each domain. However, since the global embeddings are generated based on domain-specific embeddings, the domain-shared knowledge will also be used to update domain-specific embeddings, leading to poor disentanglement.  
SAML~\cite{saml}  maps the features into global and domain-specific embeddings and incorporates a mutual unit to learn domain similarity, which is then used in fusing domain-specific knowledge.  However, it lacks effective alignment mechanisms and does not fully explore the relationships among domains.


\stitle{Dual-domain recommendation} 
Dual-domain recommendation (DDR) is another category of cross-domain recommendation that usually improves recommendation accuracy for two domains or enhances a target domain with a source domain~\cite{CRBIP, coast, DDRsurvey, CDRIB, CFAA, iihgcn, UAFCDR, CDRSur}. 
For instance, 
DisenCDR~\cite{DisenCDR} proposes two regularization terms on user embeddings based on mutual information to enforce user domain-shared representations and domain-specific representations encoding predictive and exclusive information for both domains. 
PTUPCDR~\cite{PTUPCDR} adopts meta-learning to learn user preferences in the source domain and then generates personalized models for each user in the target domain.
However, it is not clear how DDR methods can be used in MDR. 
Besides, our ED recommender and domain alignment also differ from their designs, and our domain alignment may also benefit DDR by identifying similar users/items from two domains to provide supervision signals.

\stitle{Multi-task learning} 
Multi-task learning (MTL) methods~\cite{LiLDZZ20, MaILYCCNB22, AnKKP22, LiWLS22, mln23WWW} train a single model to handle different tasks (e.g., segmentation, edge detection, classification) on the same data. Most of them follow the model disentangling architecture~\cite{MMoEMaZYCHC18}. For instance, Cross-Stitch~\cite{MisraSGH16} adopts cross-stitch units to learn the weights for linearly combining the outputs of the previous layers for each task. 
MMoE~\cite{MMoEMaZYCHC18} uses multiple networks with task-specific weights as the bottom network and is more effective when the tasks are loosely related. 
PLE~\cite{PLETangLZG20} interleaves shared model and task-specific model to improve knowledge sharing among the tasks. In particular,  each layer of the shared model connects to the next layer of all task-specific models, which in turn connects to the next layer of the shared model.
MTL has a single input data distribution but different label spaces for tasks while domains in MDR has the same label space but different data distribution~\cite{StarShengZZDDLYLZDZ21}. As shown in our experiments, for MDR, our embedding disentangling architecture outperforms model disentangling adopted by MTL methods.   




\section{Conclusion} \label{sec:con}

In this paper, we propose \name~ for multi-domain recommendation, consisting of two key designs, i.e., embedding disentangling recommender and domain alignment. In particular, the embedding disentangling recommender separates both model and embedding for capturing general and domain-specific knowledge. 
Thus, it enables the two kinds of knowledge to learn without interference. 
Domain alignment finds similar user/item pairs from different domains using random walks. The similar pairs serve as a regularization to improve knowledge sharing across domains. Extensive experiments show that \name~ outperforms state-of-the-art methods. 


\section*{Acknowledgements}
Reynold Cheng and Wentao Ning were supported by the HKU-TCL Joint Research Center for Artificial Intelligence (Project no. 200009430), the University of Hong Kong (Projects 104005858 and 10400599), the Guangdong–Hong Kong-Macau Joint Laboratory Program 2020 (Project No: 2020B1212030009), and The Hong Kong Jockey Club Charities Trust (HKJC), No. 260920140.
Bo Tang and Wentao Ning were supported by the Shenzhen Fundamental Research Program (Grant No.\ 20220815112848002) and the Guangdong Provincial Key Laboratory (Grant No. 2020B121201001).

\clearpage
\bibliographystyle{abbrv}
\balance
\bibliography{ref}


\end{document}